\documentclass{article}

\usepackage[english]{babel}
\usepackage{subcaption} 

\usepackage[letterpaper,top=2cm,bottom=2cm,left=3cm,right=3cm,marginparwidth=1.75cm]{geometry}

\usepackage{amsmath}
\usepackage{amssymb}
\usepackage{graphicx}
\usepackage[colorlinks=true, allcolors=blue]{hyperref}

\usepackage{ulem}

\newcommand{\Sref}[1]{Section~\ref{#1}}

\newcommand{\Fref}[1]{Figure~\ref{#1}}

\newcommand{\Tref}[1]{Table~\ref{#1}}
\newcommand{\tref}[1]{table~\ref{#1}}

\newcounter{daggerfootnote}
\newcommand*{\daggerfootnote}[1]{%
    \setcounter{daggerfootnote}{\value{footnote}}%
    \renewcommand*{\thefootnote}{\fnsymbol{footnote}}%
    \footnote[2]{#1}%
    \setcounter{footnote}{\value{daggerfootnote}}%
    \renewcommand*{\thefootnote}{\arabic{footnote}}%
    }

\begin{document}

\begin{center}
    \textbf{\Large Electromagnetic pressure-gradient-driven instabilities with moderate high mode numbers in tokamak plasmas}
\end{center}

\begin{center}
    Y. Narbutt$^1$\daggerfootnote{Email address for correspondence: yann.narbutt@ipp.mpg.de}, K. Aleynikova$^1$, M. Borchardt$^1$, T. Hayward-Schneider$^2$, R. Kleiber$^1$, A. {K\"onies}$^1$, A. Mishchenko$^1$, C. {N\"uhrenberg}$^1$ and E. Sánchez$^3$
\end{center}
$^1$Max Planck Institute for Plasma Physics, IPP, Wendelsteinstraße 1, 17491 Greifswald, Germany

$^2$Max Planck Institute for Plasma Physics, IPP, Boltzmannstraße 2, 85748 Garching, Germany

$^3$Laboratorio Nacional de Fusión, CIEMAT, Avda. Complutense 40, Madrid 28040, Spain

\begin{abstract}
Pressure gradient-driven instabilities are investigated in tokamak plasmas using the global gyrokinetic code EUTERPE emphasizing the role of moderate high mode numbers. As the normalized plasma pressure $\beta$ is increased, there is a well-known, sudden transition from ion-temperature-gradient (ITG) instabilities to kinetic ballooning modes (KBM), if the magnetohydrodynamic (MHD) geometry is held fixed. However, if the equilibrium field is recomputed for each value of $\beta$, so that the equilibrium is consistent with the stability calculation, the transition can disappear. In a number of cases, we are only able to find an ITG-KBM transition if inconsistent equilibria are used. In the MHD unstable regime, gyrokinetic simulations and MHD stability calculations show good agreement for moderate ratios of ion temperature gradient to density gradient and small values of the ion gyro-radius. Otherwise, non-MHD contributions are important, such as the diamagnetic stabilization and ion wave--particle resonant effects.
\end{abstract}

\section{Introduction}

In magnetic-confinement fusion, the achievable normalized plasma pressure $\beta=nT/(B^2/2\mu_0)$, the ratio of plasma pressure to magnetic pressure, is usually limited by instabilities.  When $\beta$ is small enough, these instabilities are electrostatic in nature, but become increasingly electromagnetic with increasing $\beta$. The most rapidly growing pressure-driven instabilities are those with small wavelengths perpendicular to the magnetic field. Among such small-scale instabilities, the ion-temperature-gradient (ITG) mode is usually dominant at low $\beta$ and is, at first, somewhat stabilized as $\beta$ increases \cite{aleynikovaKineticBallooningModes2018,mishchenkoGlobalGyrokineticSimulations2023,ishizawaPlasmaBetaDependence2024}. On the other hand, if $\beta$ gets large enough, pressure-driven kinetic ballooning modes (KBMs) are expected to be destabilized. The latter can be thought of as magnetohydrodynamic (MHD) ballooning modes (BMs) modified by kinetic effects such as those induced by finite-Larmor-radius orbits or particle trapping, for instance \cite{tangKineticBallooningModeTheoryGeneral1980,antonsenKineticEquationsLow1980,tsaiTheoryKineticBallooning1993,zoncaKineticTheoryLowfrequency1996,aleynikovaQuantitativeStudyKinetic2017,zoccoTrappedelectronModificationKinetic2026}. Numerical simulations of KBM linearly unstable plasmas have suggested that nonlinear heat fluxes might result in considerably high levels \cite{pueschelPropertiesHighvMicroturbulence2013,mckinneyKineticballooningmodeTurbulenceLowaveragemagneticshear2021,mulhollandFinitevTurbulenceWendelstein2024,wilmsGlobalGyrokineticSimulations2025}, making it important to understand their stability threshold. The latter usually lies below that of MHD BMs but remains less well understood. \\

KBMs have been simulated both in local flux-tube geometry \cite{aleynikovaKineticBallooningModes2018,pueschelPropertiesHighvMicroturbulence2013,mckinneyKineticballooningmodeTurbulenceLowaveragemagneticshear2021,mulhollandFinitevTurbulenceWendelstein2024,aleynikovaInfluenceMagneticConfiguration2022} and in global settings \cite{mishchenkoGyrokineticParticleincellSimulations2022,martincollarComparingElectromagneticInstabilities2020,gorlerIntercodeComparisonGyrokinetic2016,dongNonlinearSaturationKinetic2019,chenEffectsRadialElectric2023}. Nonlinear simulations of KBMs in flux-tubes usually suffer from numerical runaway of heat fluxes/conductivity, which prevents the simulations from converging in the KBM-unstable region of parameter space. Although the concept of a so-called ``non-zonal transition'' was proposed to explain this behavior \cite{pueschelPropertiesHighvMicroturbulence2013,zhangHighbetaRunawayTransitions2026,zhangZonalflowGenerationSaturation2026}, it has not fully resolved the issue. The non-zonal transition refers to the idea that zonal flows are weakened by magnetic perturbations which causes heat fluxes to diverge. Recent work by Y.~Chen et al. \cite{chenSaturationKineticBallooning2025} has found, using a global model, that including electron parallel nonlinearity in the gyrokinetic equations can help saturate nonlinear heat fluxes in the KBM regime, whereas neglecting this term leads to an unphysical blow-up of heat fluxes. Furthermore, global effects have been shown to naturally suppress the runaway problem, enabling steady-state heat fluxes in KBM simulations \cite{wilmsGlobalGyrokineticSimulations2025,mishchenkoGyrokineticParticleincellSimulations2022,chenEffectsRadialElectric2023}.\\

A critical aspect often overlooked in KBM and high-$\beta$ studies is the influence of finite plasma pressure on the magnetic equilibrium, such as the Shafranov shift. Many studies rely on simplified, analytical geometries \cite{ishizawaPlasmaBetaDependence2024,chenEffectsRadialElectric2023,ishizawaGlobalGyrokineticSimulation2019} or use inconsistent equilibria that neglect plasma pressure effects entirely \cite{gorlerIntercodeComparisonGyrokinetic2016,tronkoFirstPrinciplesGyrokinetic2019,ishidaGlobalGyrokineticNonlinear2020}. It has been demonstrated for both local and global models that ignoring these effects can lead to qualitatively different results compared to simulations with self-consistent equilibria \cite{aleynikovaKineticBallooningModes2018,mishchenkoGlobalGyrokineticSimulations2023,ishizawaPlasmaBetaDependence2024,aleynikovaQuantitativeStudyKinetic2017}. Niiro et al. \cite{niiroPlasmaBetaDependence2023} compared a simple s-$\alpha$-model and fully consistent VMEC-equilibria using a local flux-tube model and found that strong stabilization of ITG modes and strong destabilization of KBMs, commonly reported in the literature, can be artifacts of missing finite-pressure effects in the equilibrium. This finding, however, was not widely recognized in the community. Among other things, we extend these observations to a global model.\\

In this paper, we investigate pressure gradient-driven electromagnetic instabilities in tokamak plasmas, with a particular focus on the KBM for moderate high mode numbers. Previous studies have been limited by the use of local flux-tube models and/or by neglecting self-consistent equilibrium effects. Simulations focus mostly on $n=10$ as it a less explored, important limit, which is difficult to study analytically due to the lack of a good expansion parameter $n^{-1}\ll1$. Our simulations are performed with the global gyrokinetic code EUTERPE \cite{kleiberEUTERPEGlobalGyrokinetic2024}. We further compare results with the MHD code CAS3D \cite{schwabIdealMagnetohydrodynamicsGlobal1993} to assess the role of kinetic effects and to evaluate the validity of MHD modeling for the considered parameter range.\\

It is important to note that KBMs can be considered from three different perspectives. The first one is micro-instabilities where growth rates and frequencies are examined when scanning in plasma parameters, geometric quantities or specific kinetic effects. In the second one, the KBMs are related to MHD BMs because it is possible to view KBMs as kinetically modified BMs. Therefore one can correlate MHD stability with KBM stability. The third perspective is the perspective of low-frequency Alfvén modes which encompass the low frequencies of KBMs. Here, the shear Alfvén continuum couples to the acoustic continuum which gives rise to new modes such as  beta-induced Alfvén eigenmodes (BAE), electromagnetic geodesic acoustic modes and also KBMs. In this paper, we will examine all three perspectives.\\

The structure of this paper is as follows: \Sref{sec:setup} gives an overview of the simulation setup. \Sref{sec:circTok} presents results for a self-consistent tokamak for various toroidal mode numbers and $\beta$-values. The effect of finite pressure on the equilibrium is also investigated here. In \Sref{sec:circTok_CAS3D}, a comparison between CAS3D and EUTERPE is performed. Finally, \Sref{sec:discussion} discusses results of the previous chapters and concludes with a summary.\\

\section{Setup and general information}
\label{sec:setup}

To assess stability, this work employs two numerical codes: The first one is EUTERPE, a global gyrokinetic $\delta f$-PIC code that accounts for both perpendicular and parallel magnetic fluctuations, $\delta A_\parallel$ and $\delta B_\parallel$ \cite{kleiberEUTERPEGlobalGyrokinetic2024}. The second code, CAS3D, is an ideal MHD code based on the linearized MHD energy principle used to solve for growth rates and eigenfunctions \cite{schwabIdealMagnetohydrodynamicsGlobal1993}. Detailed descriptions of these codes are omitted here for conciseness but are available in the cited references. Eigenfunctions of the electrostatic potential are displayed in this work. While gyrokinetics obtains this naturally, MHD solves for the time evolution of the Lagrangian displacement vector $\xi$. This quantity can however be transformed to an electrostatic potential comparable to the one of gyrokinetics using the ideal Ohm`s law. More details on this can be found in Ref.~\cite{nuhrenbergGyrokineticSimulationsMagnetohydrodynamic2025}. All simulations presented in this work are linear.\\

The magnetic geometry investigated is a circular tokamak with a vacuum on-axis field strength $B_0=1$ T, minor radius $a=1$ m, aspect ratio $A=10$ and a safety-factor profile of $q(\rho)=1.1+0.8\rho^2$, where $\rho=r/a = \sqrt{s}$ is the normalized radial coordinate, where $s$ is the normalized toroidal flux. Simulations use a self-consistent fixed-boundary magnetic equilibrium calculated with the Variational Moments Equilibrium code (VMEC) \cite{hirshmanThreedimensionalFreeBoundary1986}. This code assumes nested flux surfaces and solves the MHD force balance $\nabla p=\vec{j}\times\vec{B}$ for a given pressure profile $p(s)$. In this context, we take the term ``self-consistent'' to  mean that the equilibrium is recalculated for every set of temperature and density profiles to ensure consistency with the resulting plasma pressure in the equilibrium and stability calculations. The safety-factor profile (and therefore also the shear profile) is fixed to the above mentioned parametrization for all calculations with VMEC. In \Sref{sec:circTok}, the local $\beta$ calculated at $s_0$ is used as the scanning parameter. From \Sref{sec:TestingMaxwellian} onward, the volume-averaged $\left<\beta\right>$ calculated by VMEC is used to compare EUTERPE results with CAS3D. In all cases, $\beta$ (and $\left<\beta\right>$) contains the total plasma pressure of both ions and electrons.\\

\begin{figure}
    \centering
    \includegraphics[width=0.5\linewidth]{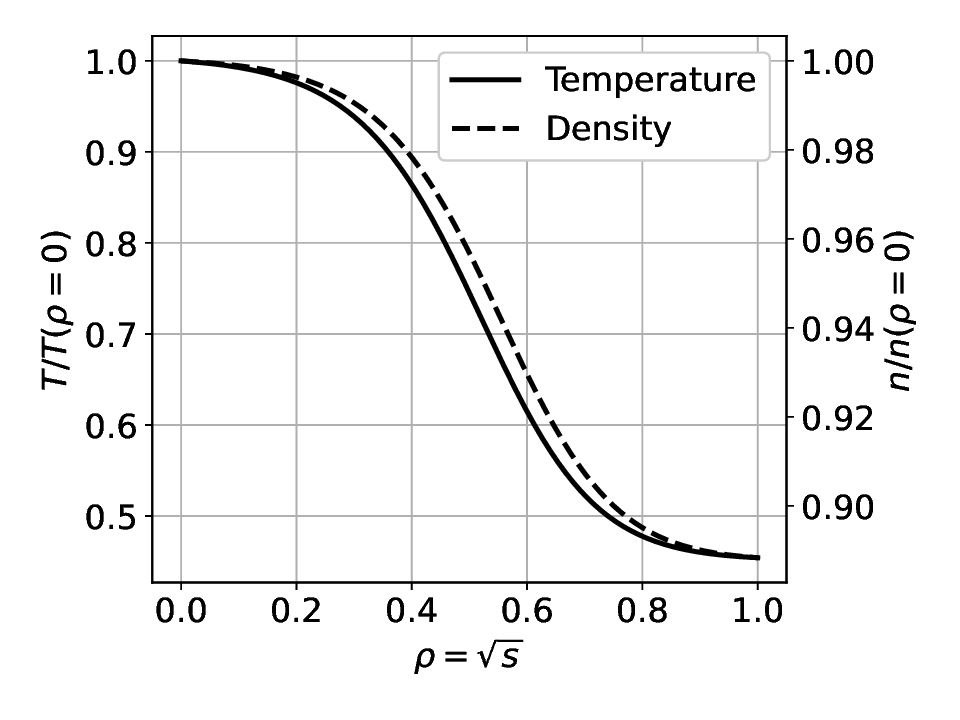}
    \label{fig:TokCirc_profiles_tanh}
    \caption{Profile examples of profile shape \eqref{eq:ProfileShape1}. Parameters used in this work are detailed in \tref{tab:TokCirc_profileParams}, with profiles and parameters explained in the text.}
    \label{fig:TokCirc_profiles}
\end{figure}

Temperature and density profiles are defined by two different expressions throughout this work. The first one is given by:
\begin{equation}
    X(s)/X_0=\exp\left(-\frac{a}{L_X}\Delta_X \tanh\left( \frac{s-s_0}{\Delta_X}\right)\right)\,,
    \label{eq:ProfileShape1}
\end{equation}
and the second by:
\begin{equation}
    \frac{1}{X}\frac{\partial X}{\partial s}=\begin{cases}
        -\frac{a}{L_X}\left(\Delta_X-\left|s-s_0\right|\right)\, & \text{if $|s-s_0|\leq\Delta_x$}\,,\\
        0 & \text{otherwise}
    \end{cases}
    \label{eq:ProfileShape2}
\end{equation}
where $X$ denotes either density or temperature, $a/L_X=(\partial X/\partial s)/X$ is the normalized gradient, $\Delta_X$ and $s_0$ are the gradient width region and position of steepest gradient, respectively. Additionally, we define $\eta=L_n/L_T$. \Tref{tab:TokCirc_profileParams} lists parameters, profile types and sections where they are used. Due to the choice of profile parameters $\eta$ remains fixed wherever a finite gradient is present. An example temperature and density profile using the first expression is displayed in \Fref{fig:TokCirc_profiles}, where $\eta=L_n/L_T$. For all simulations ions and electrons have the same temperatures and densities, $T_\mathrm{i}=T_\mathrm{e}$ and $n_i=n_e$.

When $\beta$ is varied in numerical simulations without the constraint of reproducing specific experimental data, two conventional approaches are typically employed: the first consists of keeping the temperature profile constant while scaling the density, i.e. multiplying it by various constant factors. This is the more common approach, as it maintains a constant $\rho^*=\rho_\mathrm{s}/a$, where $\rho_\mathrm{s}=\sqrt{m_\mathrm{i}k_\mathrm{B}T_\mathrm{e}}/(eB_0)$ is the sound gyro radius. It is used in \Sref{sec:circTok}. In the second approach, both plasma temperature and density are scaled simultaneously, which results in a change of both $\rho^*$ and $\beta$. These two distinct approaches can lead to different characteristics in the typical ITG-to-KBM transition as $\beta$ increases when keeping mode numbers fixed. The second approach is used for all results from \Sref{sec:TestingMaxwellian} onward, with temperature and density scaled equally. In \Sref{sec:circTok}, a reduced ion-to-electron mass ratio $m_\mathrm{i}/m_\mathrm{e}=200$ is used to reduce computational cost. The influence of this assumption on the results is evaluated in the same section. In \Sref{sec:circTok_CAS3D}, the physical mass ratio is used when comparing the gyrokinetic and MHD simulations.

The diamagnetic frequency of species $s$ is defined as
\begin{equation*}
    \omega_{\mathrm{*p}s}=\omega_{\mathrm{*n}s}+\omega_{\mathrm{*T}s}=-\frac{2 m k_\mathrm{B}T_s}{q_s a^2 B_0}\left(\frac{1}{n_s} \frac{\partial n_s}{\partial s}+\frac{1}{T_s} \frac{\partial T_s}{\partial s} \right)\,,
\end{equation*}
where $m$ is the poloidal mode number and $q_s$ the species charge. Similarly, we define
\begin{equation*}
    \omega_\mathrm{*}=\omega_\mathrm{*pi}-\omega_\mathrm{*pe}\,.
\end{equation*}
These frequencies are evaluated at the radial location of the dominant mode with poloidal mode number $m$.

\begin{table}[]
    \centering
    \begin{tabular}{|c|c|c|c|c|c|c|c|}\hline
        Section & $\Delta_X$ & $s_0$ & $a/L_T$ & $a/L_n$ & $\eta$ & Profile type & $1/\rho^*(s_0)$\\\hline\hline
        \Sref{sec:circTok} & 0.2 & 0.25 & 2.0 & 0.3 & 6.67 & \eqref{eq:ProfileShape1} & 182\\\hline
        \Sref{sec:TestingMaxwellian} & 0.2 & 0.25 & 2.0 & 0.3 & 6.67 & \eqref{eq:ProfileShape1} & 256 - 111\\\hline
        \Sref{sec:circTok_CAS3D_eta1} & 0.85 & 0.4 & 2.9 & 2.9 & 1.0 & \eqref{eq:ProfileShape2} &562 - 330\\\hline
        \Sref{sec:circTok_CAS3D_eta05} & 0.85 & 0.4 & 1.93 & 3.86 & 0.5 & \eqref{eq:ProfileShape2} &507 - 298\\\hline
    \end{tabular}
    \caption{Overview over profile types and parameters used in the corresponding result sections. The $1/\rho^*$-values have been calculated for the highest and lowest $\beta$-values simulated.}
    \label{tab:TokCirc_profileParams}
\end{table}

\section{Completeness of the model}
\label{sec:circTok}

In this section, we examine KBMs and the typical ITG-to-KBM transition with a focus on equilibrium effects. The effects of parallel currents and $\delta B_\parallel$ are also investigated. The aim is to reproduce and understand the standard picture of KBMs displayed in typical studies.

\subsection{Detailed investigation of equilibrium effects}

\begin{figure}
    \centering
    \begin{subfigure}{0.49\linewidth}
        \includegraphics[width=\linewidth]{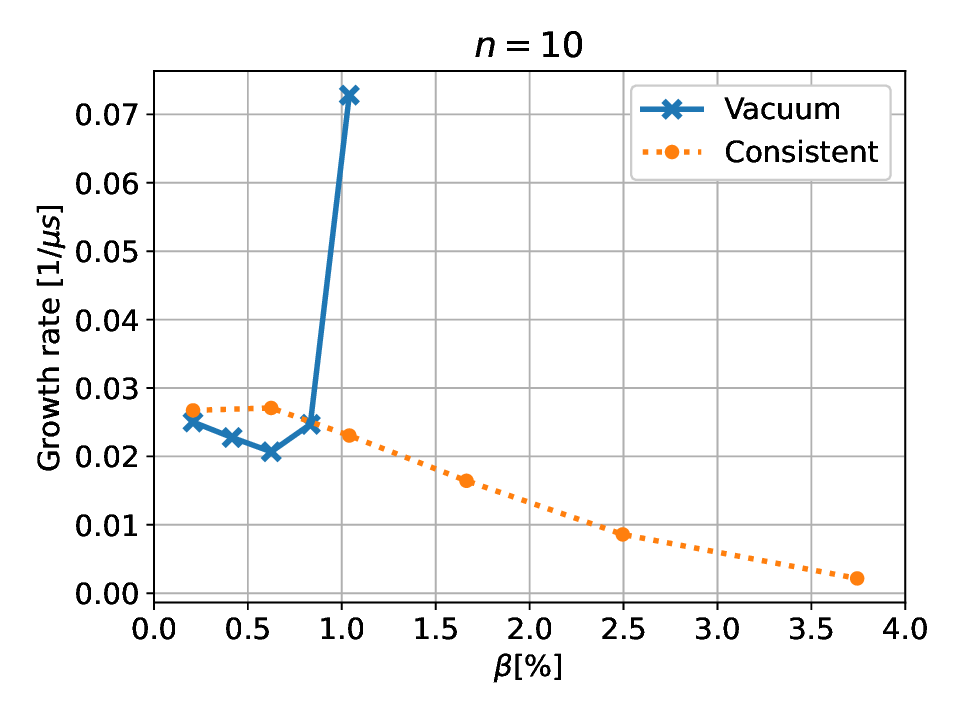}
        \caption{}
        \label{fig:TokCirc_gammaSimple}
    \end{subfigure}
    \begin{subfigure}{0.49\linewidth}
        \includegraphics[width=\linewidth]{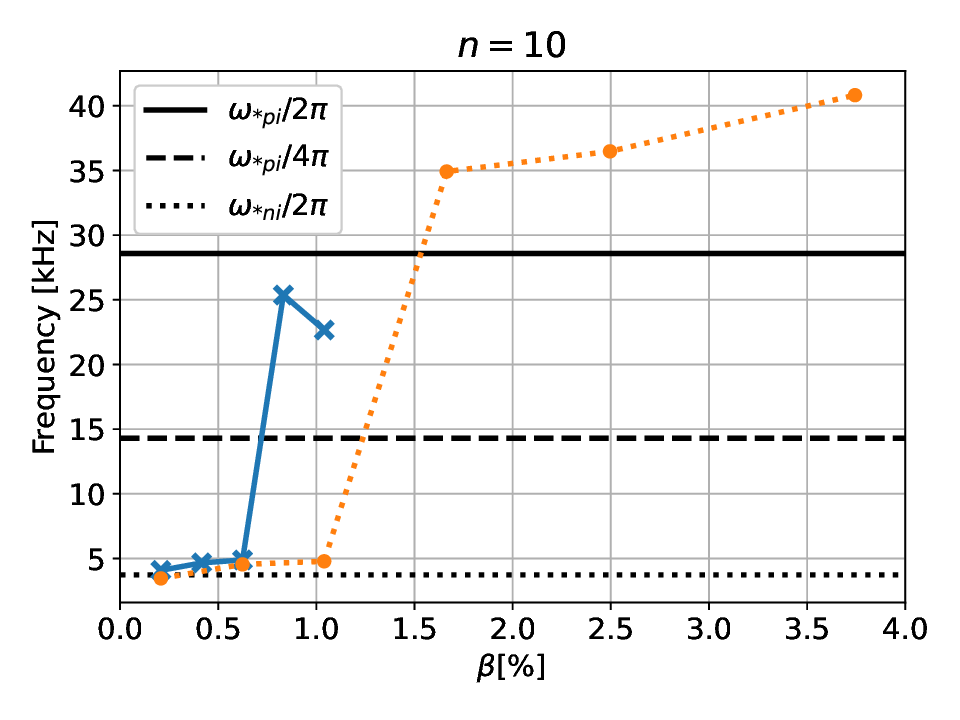}
        \caption{}
        \label{fig:TokCirc_freqSimple}
    \end{subfigure}
    \\
    \begin{subfigure}{0.49\linewidth}
        \includegraphics[width=\linewidth]{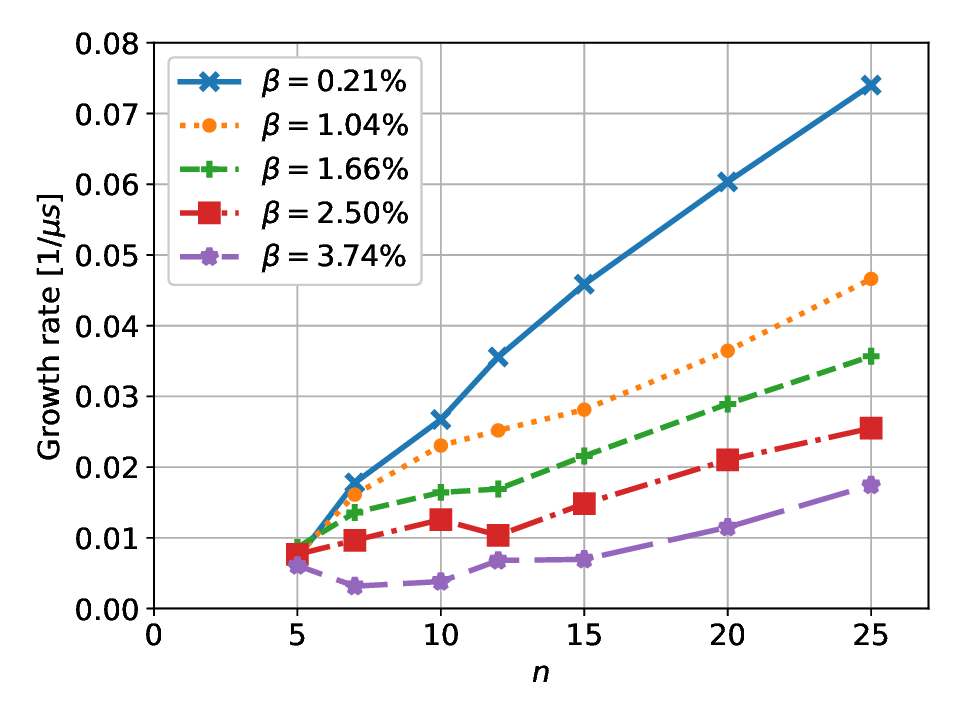}
        \caption{}
        \label{fig:TokCirc_gammaN0}
    \end{subfigure}
    \begin{subfigure}{0.49\linewidth}
        \includegraphics[width=\linewidth]{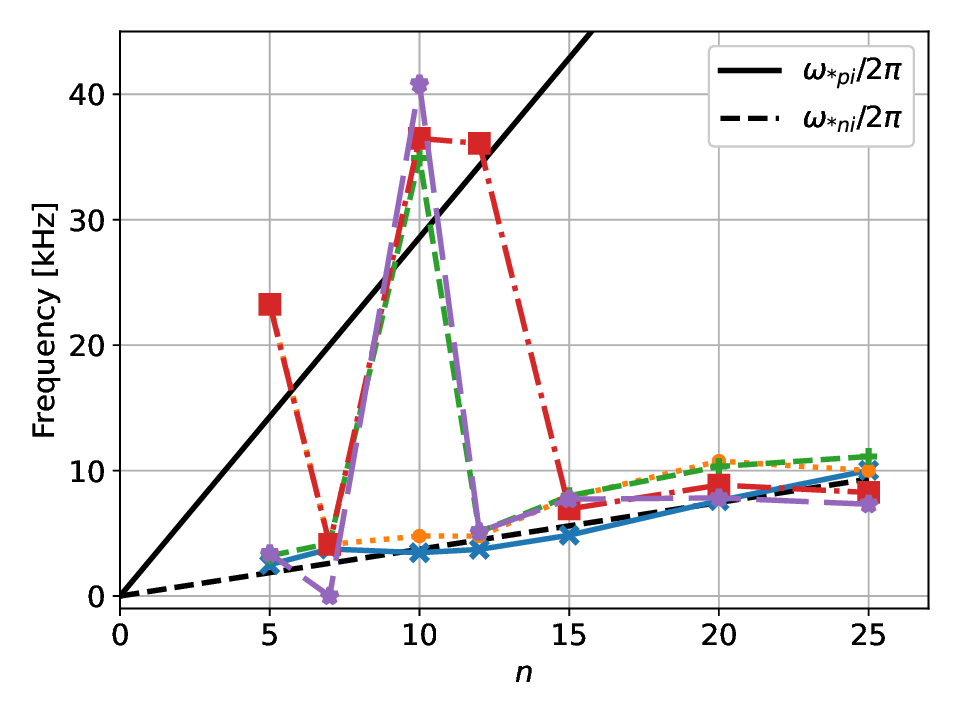}
        \caption{}
        \label{fig:TokCirc_freqN0}
    \end{subfigure}
    \caption{(a), (b) Comparison of a vacuum equilibrium with a self-consistent equilibrium for toroidal mode number $n=10$. The KBM found with vacuum geometry disappears once finite pressure effects are consistently accounted for in the geometry. A transition in frequency is still observed, but to values larger than $\omega_\mathrm{*pi}$. (c), (d) Growth rate and frequency spectra for several toroidal mode numbers $n$ and $\beta$, obtained with consistent equilibria. Growth rates tend to zero as $n\to0$. Simulations for $n=3$ were performed but showed no instability.}
    \label{fig:TokCirc_results}
\end{figure}

We vary $\beta$ by scaling the density while keeping the temperature profile fixed to achieve the desired pressures. This choice fixes the normalized gyro-radius to $1/\rho^*=182$ at mid-radius. \Fref{fig:TokCirc_gammaSimple} shows growth rates of a $\beta$-scan for toroidal mode number $n=10$. For the ``Vacuum''-labeled curve the equilibrium and instability calculations were inconsistent: a vacuum ($\beta = 0$) equilibrium was used for all values of $\beta$ used in the gyrokinetic simulation. The result is the usual ITG-to-KBM transition found by many other authors. The scan first shows a slight decrease in growth rate with increasing $\beta$, followed by a characteristically rapid increase above a threshold $\beta\approx0.8\%$. Since these simulations were performed with a vacuum equilibrium, all finite-pressure effects on the magnetic geometry, such as the Shafranov shift, are neglected. It is however well known that these are stabilizing to curvature-driven instabilities \cite{connorEffectPressureGradients1983a}. In a second set of gyrokinetic simulations, we therefore included these effects by recalculating the VMEC equilibrium for each value of $\beta$ (i.e. for every pair of temperature and density profiles). For the resulting  equilibria, the curve labeled ``Consistent'' in \Fref{fig:TokCirc_gammaSimple} is obtained. Note that the growth rate now decreases as $\beta$ increases instead of increasing beyond a certain critical $\beta$.\\

The frequencies corresponding to these growth rates are shown in \Fref{fig:TokCirc_freqSimple}. For the ITG mode in both equilibrium types, i.e. vacuum and consistent, the frequencies are close to the diamagnetic density frequency $\omega_\mathrm{*n}$. The frequencies for the vacuum case exhibit the typical jump, after the transition in growth rates, to values with $\omega_\mathrm{*pi}/2<\omega<\omega_\mathrm{*pi}$. While for the consistent case the frequencies also exhibit a jump, this is not reflected in the growth rates, and the values after the jump are larger than $\omega_\mathrm{*pi}$.\\

The combined behavior of growth rates and frequencies (growth rates strongly increase with $\beta$ and the frequency jumps to $\omega_\mathrm{*pi}/2\lesssim\omega\lesssim\omega_\mathrm{*pi}$) in the vacuum case is the standard signature used to identify KBMs used in many previous studies \cite{aleynikovaKineticBallooningModes2018,mulhollandFinitevTurbulenceWendelstein2024,gorlerIntercodeComparisonGyrokinetic2016,coleTokamakITGKBMTransition2021}. We therefore conclude that, when the equilibrium is treated self-consistently, the ITG-to-KBM transition disappears here as the standard criteria are not fulfilled.\\

To investigate the behavior for different mode numbers, and to exclude the possibility of missing the KBM by not choosing the right mode number, a scan (using consistent equilibria) in the toroidal mode number $n$ was performed (\Fref{fig:TokCirc_gammaN0}). The growth rates drop with increasing $\beta$ and generally decrease as $n$ decreases. For $n=3$ no instability could be found anymore. The corresponding frequencies (\Fref{fig:TokCirc_freqN0}) either follow $\omega_\mathrm{*ni}$, indicating an ITG mode, or, when exhibiting a jump to larger frequencies, values are larger than $\omega_\mathrm{*pi}$ which is atypical for KBMs. Thus, according to the standard criteria, for a broad range of toroidal mode numbers and $\beta$ no KBMs could be found.\\

\begin{figure}
    \centering
    \includegraphics[width=0.5\linewidth]{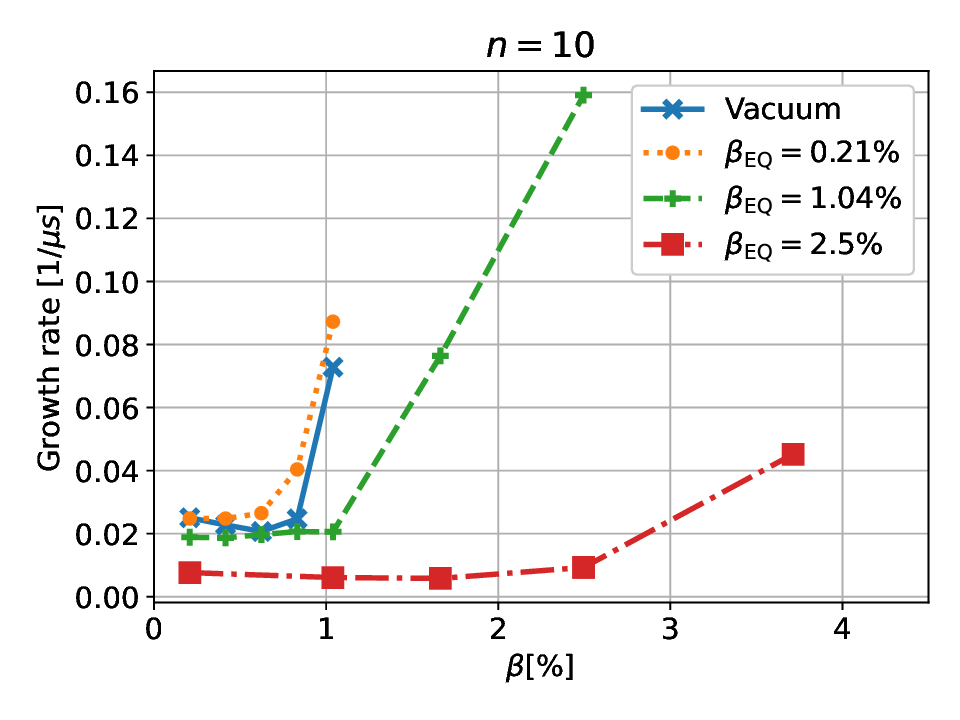}
    \caption{The effect of equilibrium consistency by comparing growth-rate trends as $\beta$ is varied in the simulations, for several equilibria: a vacuum case and three fixed equilibrium pressure cases at $\beta_\mathrm{EQ}=0.21\%$, $1.04\%$ and $2.5\%$.}
    \label{fig:TokCirc_gammaEquils}
\end{figure}

We have observed large differences in results between the case of treating the equilibrium self-consistent or using a vacuum equilibrium. To gain more insight, we now vary the equilibrium only partially. In \Fref{fig:TokCirc_gammaEquils}, three additional $\beta$-scans are shown. Instead of fixing the equilibrium at zero pressure (the vacuum case), the equilibria are now held at a finite pressure with $\beta_\mathrm{EQ}=0.21\%$, $1.04\%$ and $2.5\%$. We will refer to these as low-$\beta$, medium-$\beta$, and high-$\beta$ equilibria, respectively. Here, $\beta_\mathrm{EQ}$ refers to the local $\beta$, calculated at $s=0.5$, that is used in the equilibrium calculation of VMEC. For each scan the equilibrium configuration (and thus $\beta_\mathrm{EQ}$ as well) is kept fixed while $\beta$ is varied by changing profiles in the simulation.\\

Simulation results for the low-$\beta$ equilibrium are almost identical to those obtained with the vacuum equilibrium shown earlier. This is no surprise because the low-$\beta$ equilibrium is quite close to the vacuum equilibrium. More pronounced effects appear for the medium-$\beta$ and the high-$\beta$ equilibria: 
\begin{enumerate}
    \item No appreciable stabilization of the ITG mode is observed when varying $\beta$ but keeping the equilibrium $\beta_\mathrm{EQ}$ fixed. 
    \item Growth rates at all $\beta$-values decrease proportionally to $\beta_\mathrm{EQ}$, i.e. the higher $\beta_\mathrm{EQ}$ is, the lower the growth rates tend to be over the entire curve.
    \item The critical $\beta$ for KBMs, $\beta_\mathrm{crit}^\mathrm{KBM}$, shifts to significantly larger values as $\beta_\mathrm{EQ}$ increases.
    \item For $\beta>\beta_\mathrm{crit}^\mathrm{KBM}$ the growth rate increases less rapidly for larger $\beta_\mathrm{EQ}$. 
\end{enumerate}
These stabilizing trends become even clearer when the equilibrium is varied consistently with $\beta=\beta_\mathrm{EQ}$ while all electromagnetic terms are switched off in the gyrokinetic equations. Though not shown here, the resulting electrostatic ITG mode shows reduced growth rates as $\beta_\mathrm{EQ}$ increases, further underlining the influence of equilibrium effects.

The stabilization occurs thanks to changes in the geometry of the magnetic field as $\beta$ increases. In the present configuration, the Shafranov shift reaches $20\%$ of the minor radius at $\beta\approx2.5\%$, strongly compressing the outboard flux surfaces. However, the pressure gradient also modifies other aspects of the magnetic geometry, such as the local safety-factor profile and the Pfirsch-Schlüter current. It is therefore difficult to pinpoint the exact mechanism responsible for the observed stabilization. 

In summary, in the parameter range considered here, finite-pressure equilibrium effects raise $\beta_\mathrm{crit}^\mathrm{KBM}$ and lower the growth rates of the electromagnetic modes found.\\

\subsection{The influence of parallel current}
\label{sec:TestingMaxwellian}
All gyrokinetic simulations we have presented so far have omitted the effect of equilibrium currents. These currents are often suppressed to avoid current-driven instabilities such as tearing or kink modes, but are intrinsic to MHD and therefore belong to a fully self-consistent treatment. In EUTERPE the electron background distribution function, $F_{0,e}$ can be specified as a shifted Maxwellian distribution, which carries a parallel plasma current. The shifted Maxwellian is
\begin{equation}
    F_{0,e}\propto\exp\left(-(v_\parallel-u_{\rm e})^2/(2 v_{\rm th, e}^2)\right)\,.
\end{equation}
where $v_{\rm th, e}$ is the electron thermal velocity and $v_\parallel$ the parallel velocity. The shift velocity $u_e=u_e(\rho,\vartheta,\varphi)$ is directly related to the parallel plasma current (including the Pfirsch-Schl\"uter contribution). This shift depends on the radial coordinate $\rho$, the poloidal angle $\vartheta$, and, in stellarators, the toroidal $\varphi$. 

We now compare simulations using a shifted Maxwellian and a centered Maxwellian (i.e. $u_e = 0$) equilibrium distribution function. For the ions, the distribution function is always the standard centered Maxwellian.
\begin{figure}
    \centering
    \includegraphics[width=0.5\linewidth]{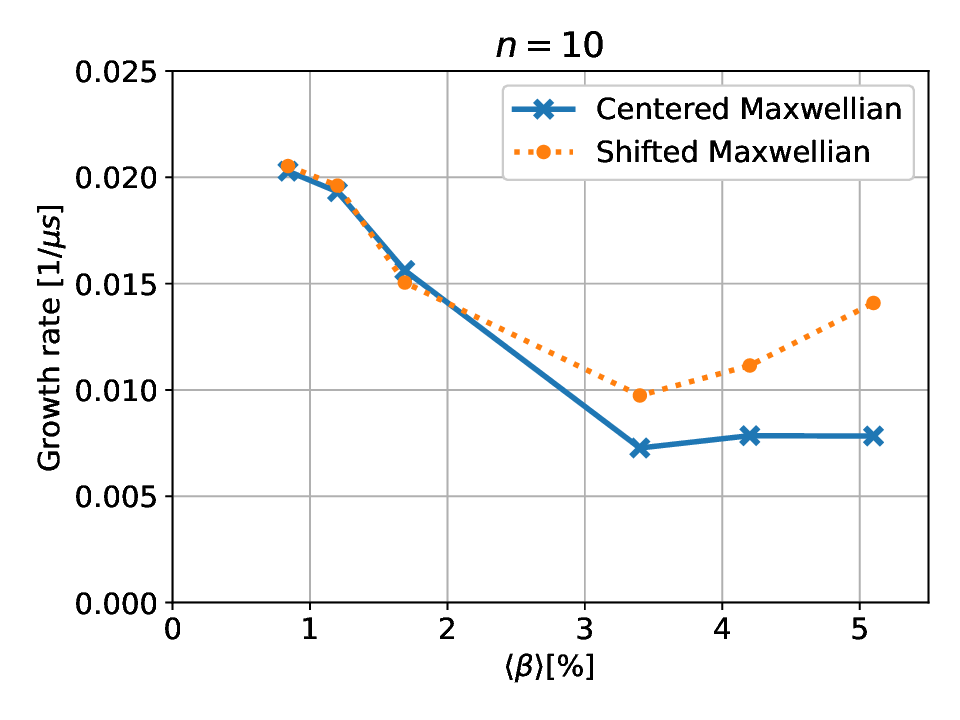}
    \caption{Testing the effect of a centered Maxwellian versus a shifted Maxwellian electron distribution on growth rates trends in simulations that use self-consistent equilibria.}
    \label{fig:TokCircCAS3D_gamma_Maxwellian}
\end{figure}

\Fref{fig:TokCircCAS3D_gamma_Maxwellian} shows a $\beta$-scan that compares results for the two electron distribution functions while the equilibrium is kept self-consistent. At low $\left<\beta\right>$, growth rates obtained with the centered Maxwellian and shifted Maxwellian are nearly identical. For $\left<\beta\right>>3.4\%$ the two cases diverge: with the shifted Maxwellian the growth rates continue to increase with $\left<\beta\right>$, whereas with the centered Maxwellian growth rates stay approximately constant past $\left<\beta\right>\approx3.4\%$. Consequently, all simulations in \Sref{sec:circTok_CAS3D} and thererafter include the parallel equilibrium current by using a shifted Maxwellian. 

\subsection{Mass ratio dependence}
All simulations in \Sref{sec:circTok} used a reduced ion-to-electron mass ratio $m_\mathrm{i}/m_\mathrm{e}=200$. When the realistic mass ratio $m_\mathrm{i}/m_\mathrm{e}=1836$ is employed for the simulations in \Sref{sec:TestingMaxwellian}, the absolute growth rates are significantly lower, although the qualitative dependence on $\left<\beta\right>$ is unchanged. In \Sref{sec:circTok_CAS3D} we adopt the realistic mass ratio to enable a more direct comparison with MHD results.\\

\subsection{Examining the effect of $\delta B_\parallel$}
\begin{figure}
    \centering
    \includegraphics[width=0.5\linewidth]{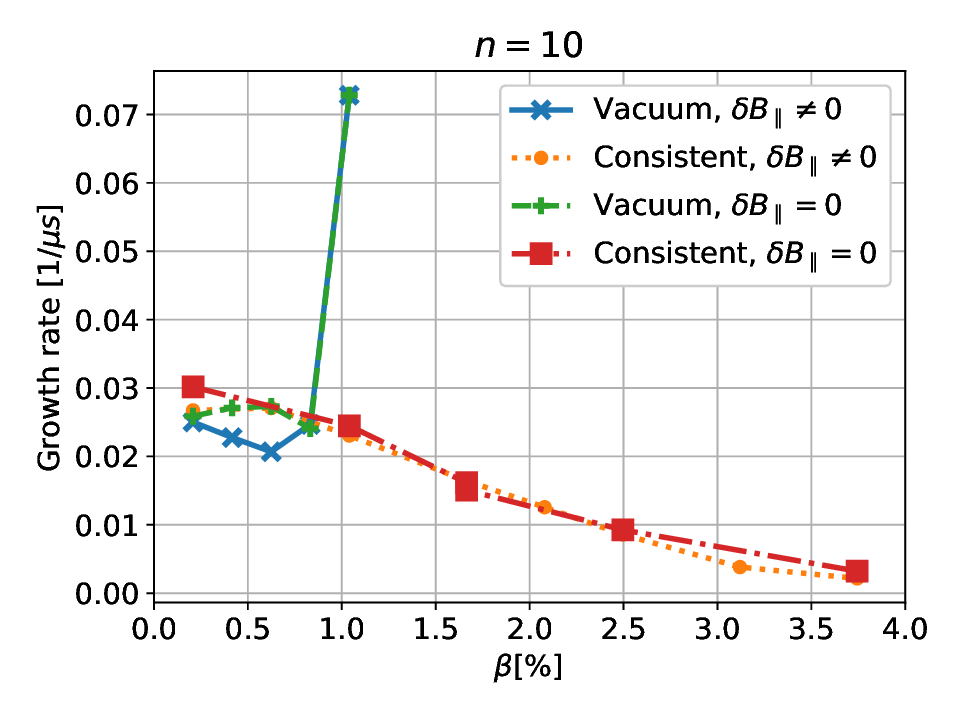}
    \caption{Effect of omitting the parallel magnetic perturbation $\delta B_\parallel$ on the results from \Fref{fig:TokCirc_results} for $n=10$, comparing the vacuum and the self-consistent equilibria.}
    \label{fig:TokCirc_gammaBpar}
\end{figure}

Including the parallel magnetic perturbation $\delta B_\parallel$ also does not destabilize the KBMs, which is in contrast to Refs.~\cite{aleynikovaKineticBallooningModes2018,kennedyImportanceParallelMagneticfield2024}. In \Fref{fig:TokCirc_gammaBpar} the effect of setting $\delta B_\parallel=0$ is shown, for both the vacuum and consistent $n=10$ simulations from \Fref{fig:TokCirc_results}. No significant change in growth rates or frequencies is observed for any of the cases considered.\\

\section{Comparison of the gyrokinetic model with the MHD model}
\label{sec:circTok_CAS3D}

In the previous section, we found that the KBMs are not as easily destabilized as suggested by most of the existing literature. Nevertheless, it is well known experimentally that, at some point, the plasma should become unstable once the MHD stability limit is exceeded. We therefore examine how the gyrokinetic plasma behaves for different parameters and plasma profiles, and compare with ideal-MHD calculations. The simulation setup described in \Sref{sec:setup} applies to the gyrokinetic and the ideal-MHD calculations, using EUTERPE and CAS3D, respectively.\\

\subsection{\texorpdfstring{$\eta=6.67$}{\eta=6.67}-case}
\label{sec:circTok_CAS3D_eta6}

\begin{figure}
    \centering
    \begin{subfigure}{0.49\linewidth}
        \includegraphics[width=\linewidth]{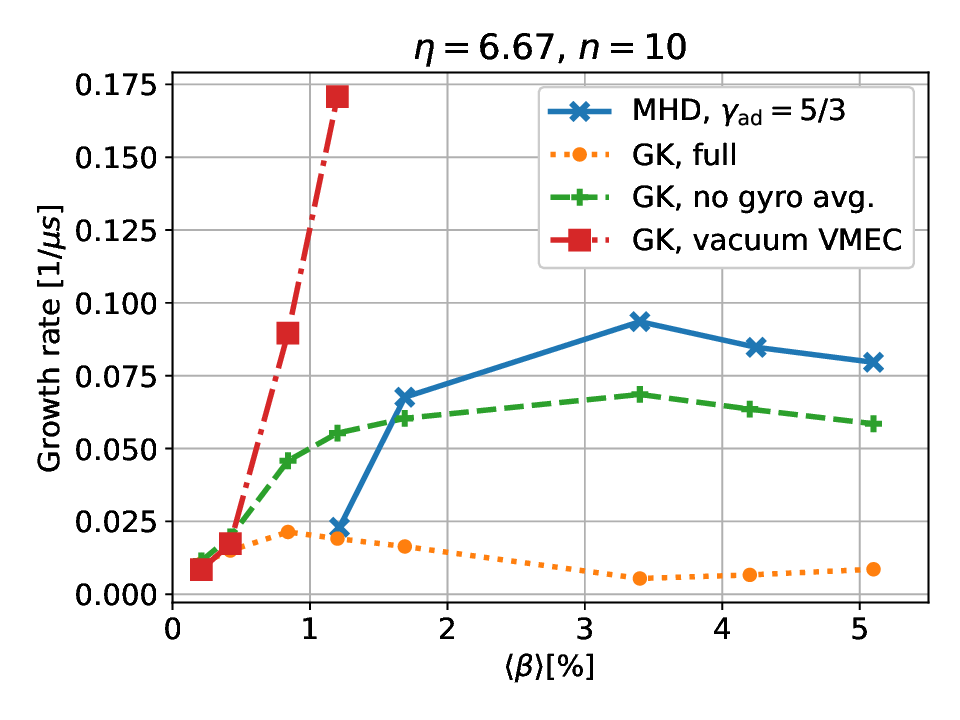}
        \caption{}
        \label{fig:TokCircCAS3D_gamma}
    \end{subfigure}
    \begin{subfigure}{0.49\linewidth}
        \includegraphics[width=\linewidth]{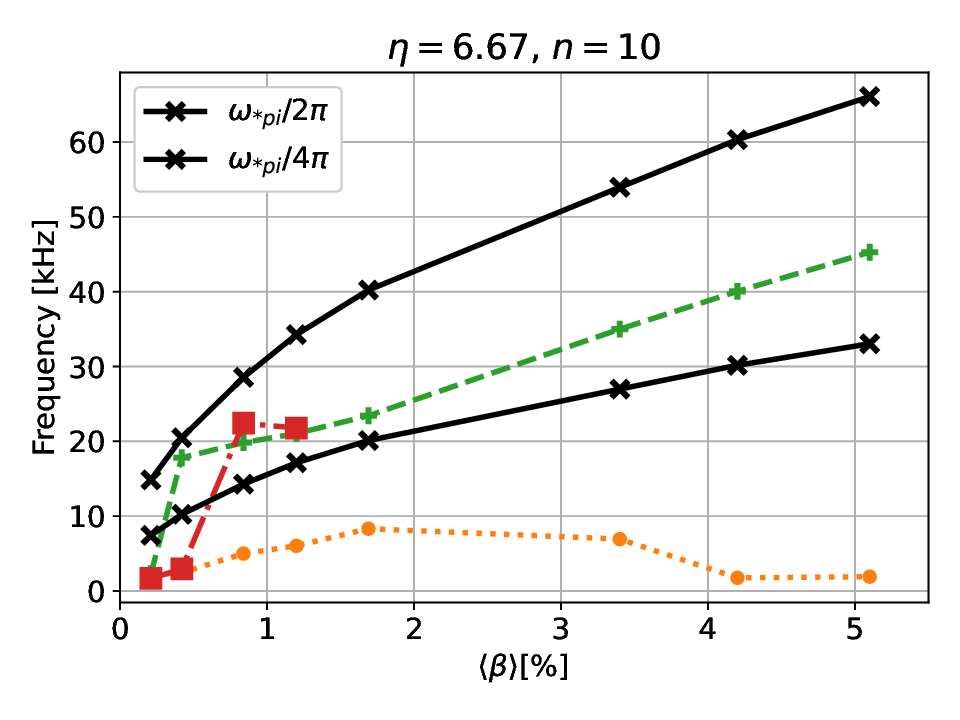}
        \caption{}
        \label{fig:TokCircCAS3D_freq}
    \end{subfigure}
    \caption{Comparison of gyrokinetic (EUTERPE) and MHD (CAS3D) results for $\eta=6.67$. For the ``vacuum''-curve, the corresponding volume averaged $\beta$ is plotted despite simulations executed in vacuum ($\beta_\mathrm{EQ}=0$). Temperature and density profiles are identical for all curves. The legend of figure (a) also applies to figure (b). As a reminder: the diamagnetic frequencies are calculated at the radial position of the dominant mode number $m$.}
    \label{fig:TokCircCAS3D_results}
\end{figure}

\Fref{fig:TokCircCAS3D_results} shows the results obtained with both MHD and gyrokinetic simulations. First, a $\beta$-scan is performed with equilibrium pressure fixed at $\beta_\mathrm{EQ}=0$ in order to observe the ITG-to-KBM transition. In \Fref{fig:TokCircCAS3D_gamma}, the growth rate curve initially shows a slight rise in the ITG regime, followed by a sharp increase once the plasma enters the KBM regime. When the growth rates sharply increase, the frequency makes a sudden transition to larger values. This behavior is typically associated with the ITG-to-KBM transition \cite{aleynikovaKineticBallooningModes2018,mulhollandFinitevTurbulenceWendelstein2024,gorlerIntercodeComparisonGyrokinetic2016,coleTokamakITGKBMTransition2021}. The transition looks somewhat different from that in \Fref{fig:TokCirc_gammaSimple}, where the growth rate  first decreases with increasing $\left<\beta\right>$ before shooting up. This difference is due to the fact that, in \Fref{fig:TokCircCAS3D_gamma}, both density and temperature are varied rather than only the density.\\

When a self-consistent $\left<\beta\right>$-scan is performed, the growth rate also shows a slight increase at low $\left<\beta\right>$, but decrease as equilibrium-pressure effects become important. Only beyond $\left<\beta\right>\approx3.4\%$ do the growth rates rise again, but only slightly. In contrast, the growth rates obtained with CAS3D show an ideal MHD instability starting at $\left<\beta\right>\approx1\%$ with a maximum at $\left<\beta\right>=3.4\%$.\\

Obviously, the MHD and gyrokinetic results do not match. It is likely that kinetic effects dominate the gyrokinetic instabilities and overshadow the MHD physics. To demonstrate this, we deactivate the gyro-average in the gyrokinetic equation for the ions (the respective weight-evolution equation solved by EUTERPE). The gyro averaging operator in Kleiber et al. \cite{kleiberEUTERPEGlobalGyrokinetic2024} is replaced by the identity (which affects the field equations and equations of motion). This does not deactivate FLR-effects entirely since gyrokinetic polarization is still included but it affects (eliminates) the diamagnetic corrections (diamagnetic stabilization). The resulting curve, also shown in \Fref{fig:TokCircCAS3D_gamma}, now agrees reasonably well with the one for MHD growth rates, except that the gyrokinetic model is already unstable at $\left<\beta\right>$ values below the MHD threshold of $\left<\beta\right>\approx1\%$. The frequency plot in \Fref{fig:TokCircCAS3D_freq} shows a transition at very low $\left<\beta\right>$. For larger $\left<\beta\right>$, the frequencies lie between the diamagnetic frequency $\omega_\mathrm{*pi}$ and half that value. We conclude that EUTERPE now recovers the BM found by MHD that is otherwise strongly modified by the kinetic (gyro-average) effects.\\

\begin{figure}
    \centering
    \begin{subfigure}{0.49\linewidth}
        \includegraphics[width=\linewidth]{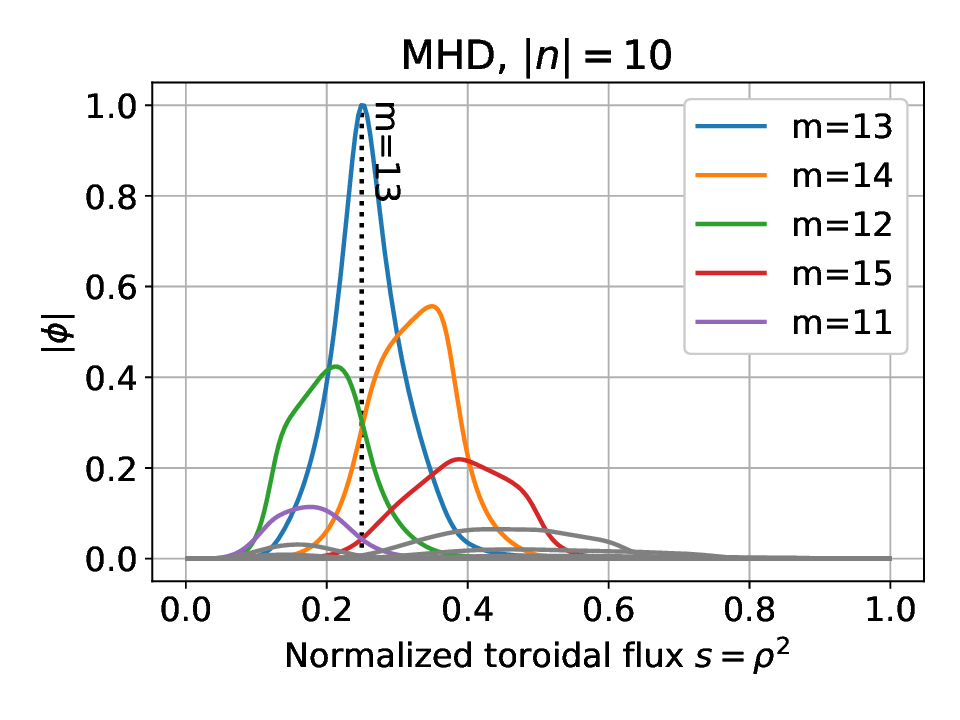}
        \caption{}
        \label{fig:TokCircCAS3D_mode}
    \end{subfigure}
    \begin{subfigure}{0.49\linewidth}
        \includegraphics[width=\linewidth]{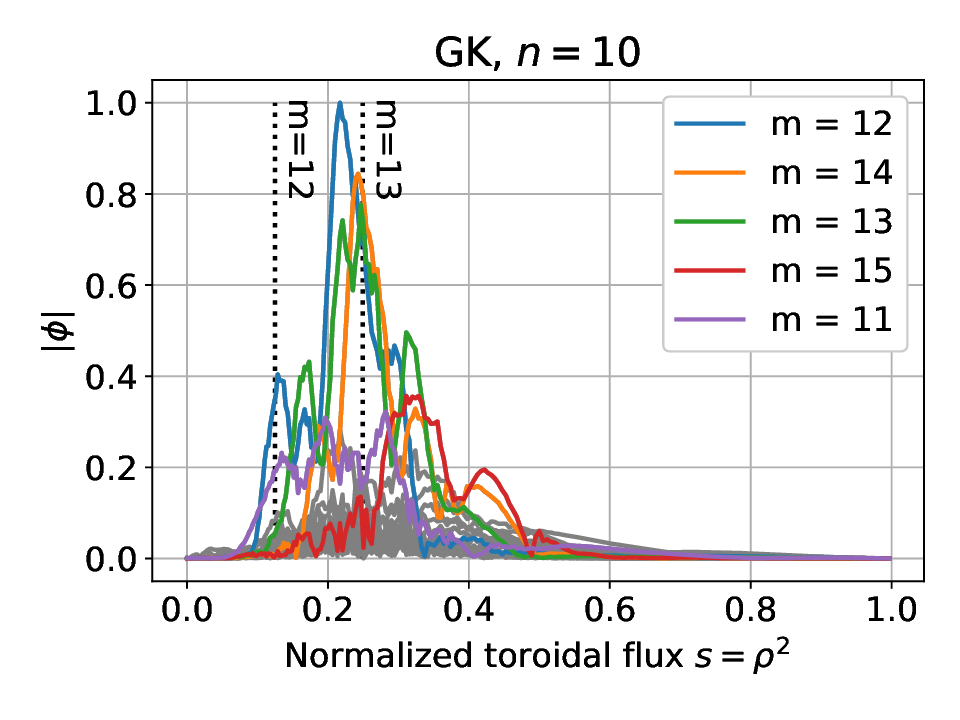}
        \caption{}
        \label{fig:TokCircCAS3D_EUmode}
    \end{subfigure}
    \hfill
    \begin{subfigure}{0.49\linewidth}
        \includegraphics[width=\linewidth]{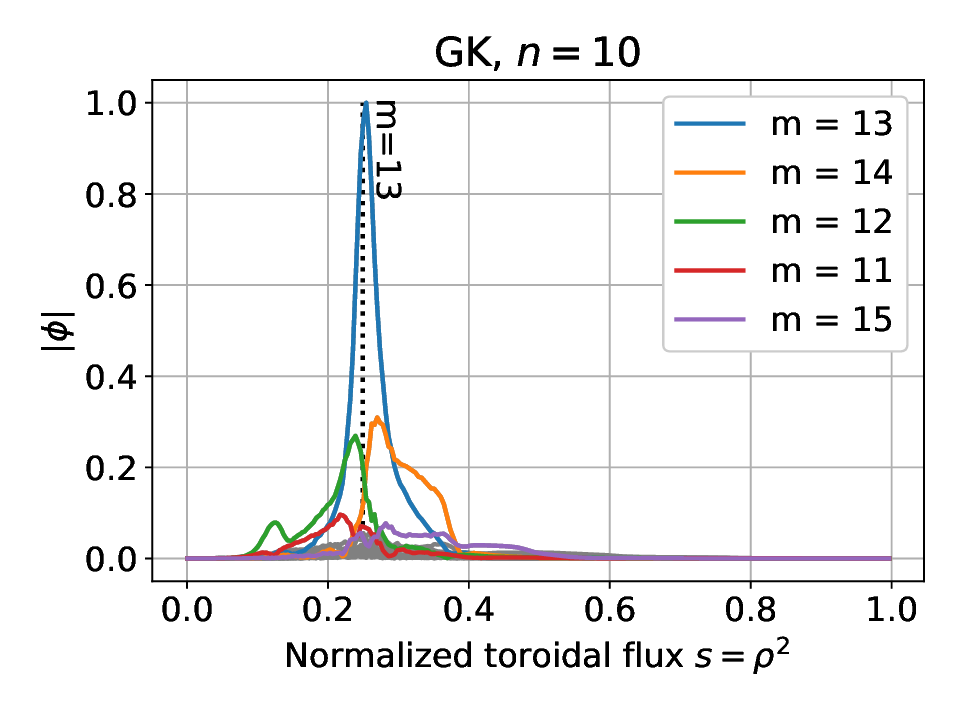}
        \caption{}
        \label{fig:TokCircCAS3D_EUmode_noFLR}
    \end{subfigure}
    \begin{subfigure}{0.49\linewidth}
        \includegraphics[width=\linewidth]{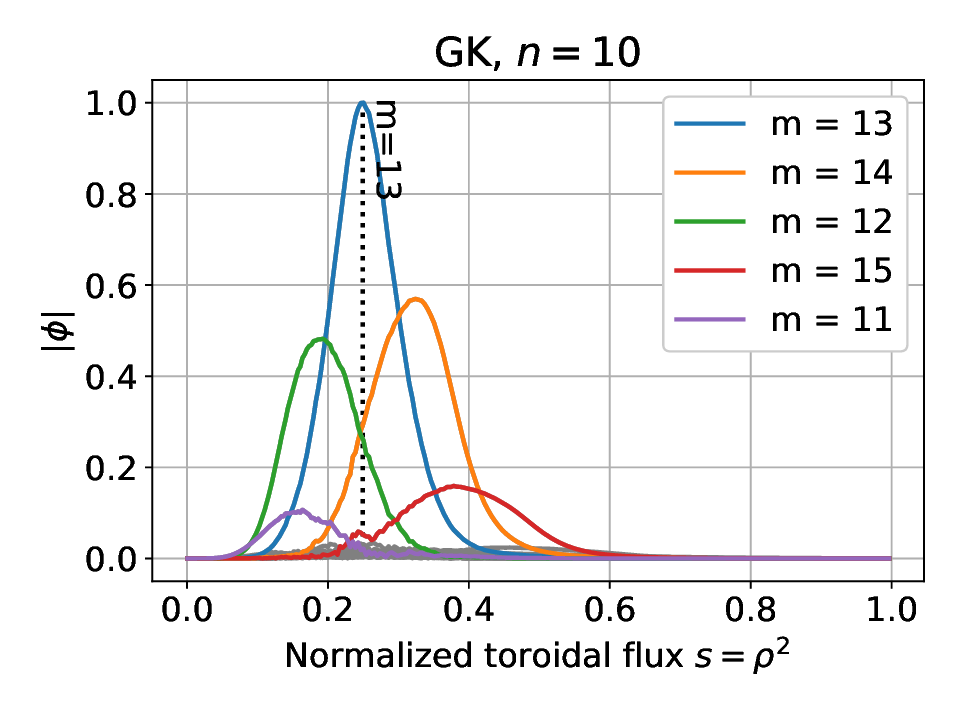}
        \caption{}
        \label{fig:TokCircCAS3D_EUmode_vac}
    \end{subfigure}
    \caption{Mode structures of electrostatic potential $\phi$ for $n=10$ and $\eta=6.67$ at $\left<\beta\right>=1.69\%$ for MHD (a), full gyrokinetics (b), gyrokinetics without gyro-average (c) and full gyrokinetics with vacuum geometry (d). Vertical dotted lines mark the radial locations of the q-profile resonances. The electrostatic potential of the MHD result is obtained by transforming the displacement $\xi$.}
    \label{fig:TokCircCAS3D_modes}
\end{figure}

\Fref{fig:TokCircCAS3D_modes} displays the mode structures of the electrostatic potential for $\left<\beta\right>=1.69\%$. The CAS3D result (\Fref{fig:TokCircCAS3D_mode}) shows a typical MHD structure. By contrast, the full gyrokinetic simulation (\Fref{fig:TokCircCAS3D_EUmode}) shows a markedly different structure. When the gyro-average is switched off (\Fref{fig:TokCircCAS3D_EUmode_noFLR}) the mode pattern becomes much more similar to the MHD results, indicating that the gyro-radius creates a strong radial coupling between the neighboring flux surfaces that distorts the MHD eigenfunction. Despite the improved agreement, some differences remain, likely because further kinetic effects (guiding center orbits, wave-particle resonances, etc.) are still present. The closest match to the MHD mode is obtained with the vacuum-geometry simulation (\Fref{fig:TokCircCAS3D_EUmode_vac}) where the gyro-average is retained but the equilibrium pressure is set to zero. This demonstrates that equilibrium effects, most notably the Shafranov shift compressing outboard flux surfaces, act together with the gyro-average to suppress the MHD ballooning instability.\\
In these figures, the dominant resonances with the $q$-profile are indicated by dotted vertical lines. When the gyrokinetic results resemble the MHD result, the dominant poloidal mode numbers coincide and are radially located at their respective resonant surface. In the full gyrokinetic case this is not true: the dominant mode numbers differ from the MHD result and the structures can be shifted away radially from their respective $q$-resonance. In particular, the dominant $m=12$ component is located at the steepest gradient ($s=0.25$). These observations underline the importance of the global gyrokinetic approach to low-frequency instabilities in toroidal plasmas.\\

\subsection{\texorpdfstring{$\eta=1$}{\eta=1}-case}
\label{sec:circTok_CAS3D_eta1}

We now examine how the details of the plasma profiles influence the results presented in the previous section. The temperature and density gradients are now chosen such that $\eta=1$ but the pressure profile remains the same. For certain stellarator equilibria, Ref.~\cite{nuhrenbergGyrokineticSimulationsMagnetohydrodynamic2025} reported excellent agreement between gyrokinetic and MHD simulations for $\eta=1$.\\

\begin{figure}
    \centering
    \begin{subfigure}{0.49\linewidth}
        \includegraphics[width=\linewidth]{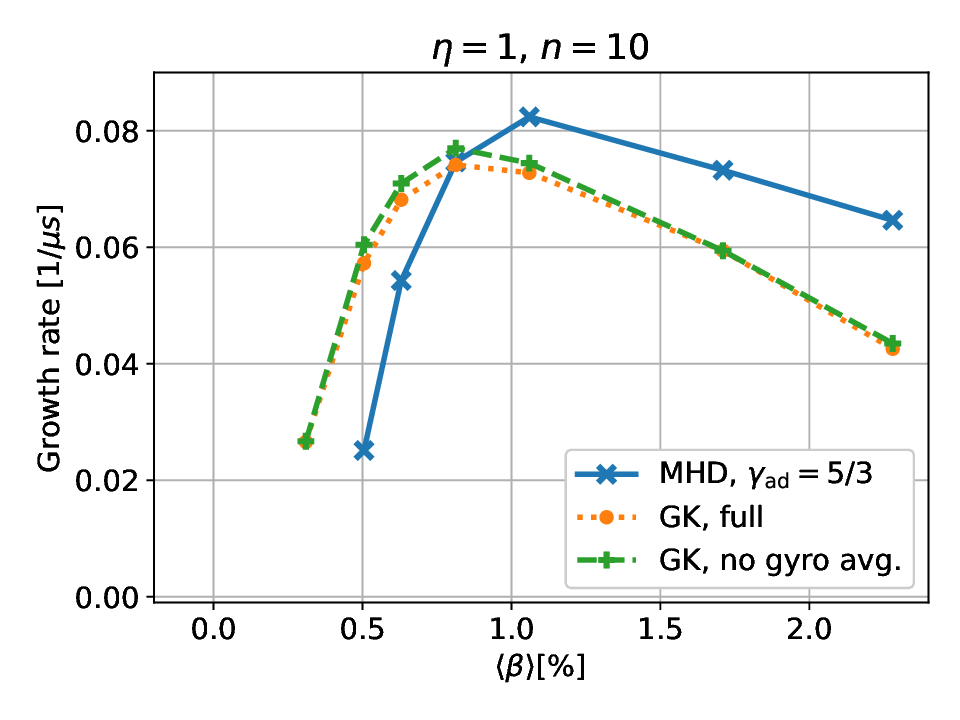}
        \caption{}
        \label{fig:TokCircCAS3D_eta1_gamma}
    \end{subfigure}
    \begin{subfigure}{0.49\linewidth}
        \includegraphics[width=\linewidth]{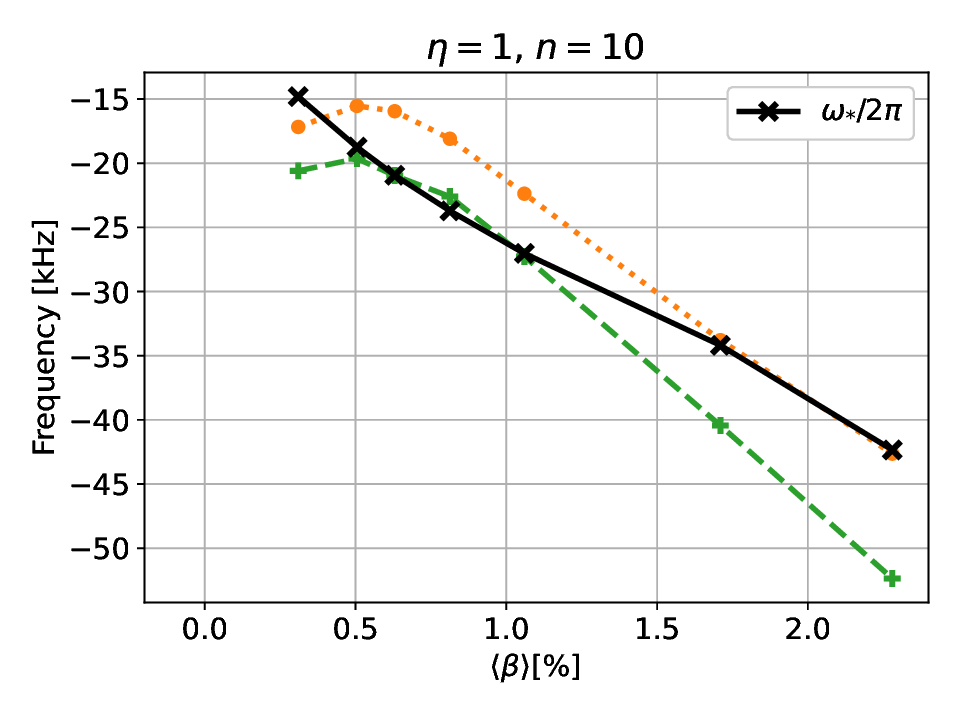}
        \caption{}
        \label{fig:TokCircCAS3D_eta1_freq}
    \end{subfigure}
    \caption{Comparison of (a) growth rates and (b) frequencies for $\eta=1$ between EUTERPE and CAS3D. A second CAS3D $\left<\beta\right>$-scan with $\gamma_\mathrm{ad}=10^{-5}$ results in slightly up-shifted growth rates, but is not displayed here for visual clarity. The legend from figure (a) also belongs to figure (b).}
    \label{fig:TokCircCAS3D_eta1_results}
\end{figure}

Simulation results from EUTERPE and CAS3D are shown in \Fref{fig:TokCircCAS3D_eta1_results}. The CAS3D growth rates in \Fref{fig:TokCircCAS3D_eta1_gamma} follow the same trend as in \Fref{fig:TokCircCAS3D_gamma}: the plasma becomes unstable for $\left<\beta\right>>0.3$. The growth rates increase until $\left<\beta\right>\approx1\%$ and then decrease again. EUTERPE also predicts the onset of an instability, but it appears at a lower $\left<\beta\right>$. Apart from the small relative shift between the two curves, the overall behavior is very similar. Unlike the $\eta=6.67$ case, turning off the gyro-average does not produce any noticeable effect. We conclude that, in this regime, the instability is more MHD-fluid-like and kinetic effects are less important. A noteworthy difference from the $\eta=6.67$ case is that the modes obtained with EUTERPE now rotate in the electron-diamagnetic direction, instead of the ion diamagnetic direction observed previously. The frequencies are very close to $\omega_\mathrm{*}$, indicating that a different physics regime is observed (this is discussed further in \Sref{sec:discussion}).\\

\begin{figure}
    \centering
    \begin{subfigure}{0.32\linewidth}
        \includegraphics[width=\linewidth]{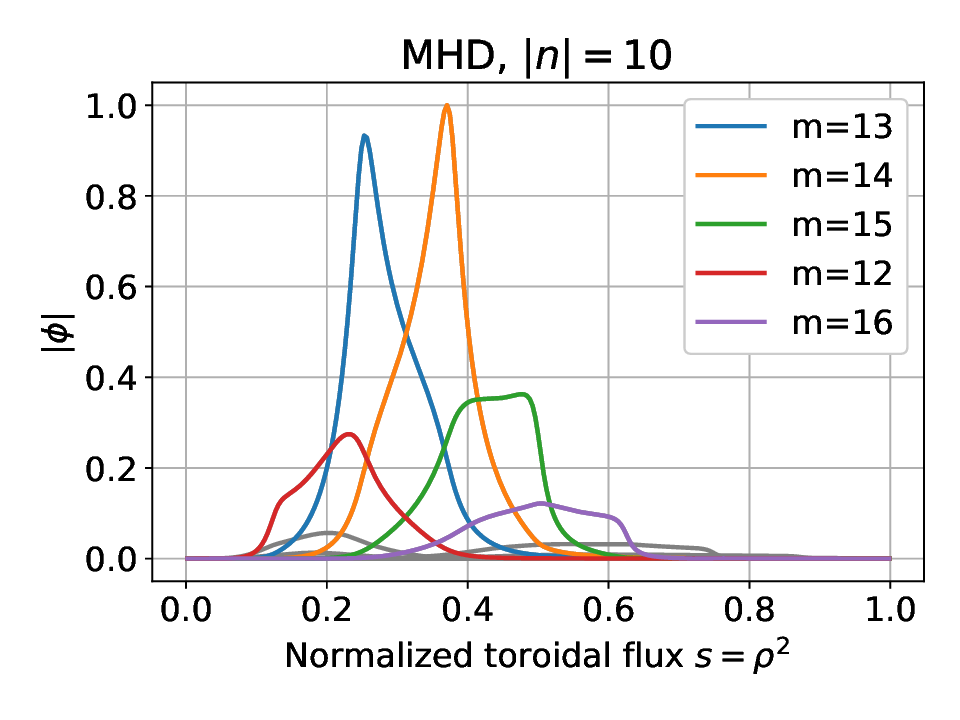}
        \caption{}
    \end{subfigure}
    \begin{subfigure}{0.32\linewidth}
        \includegraphics[width=\linewidth]{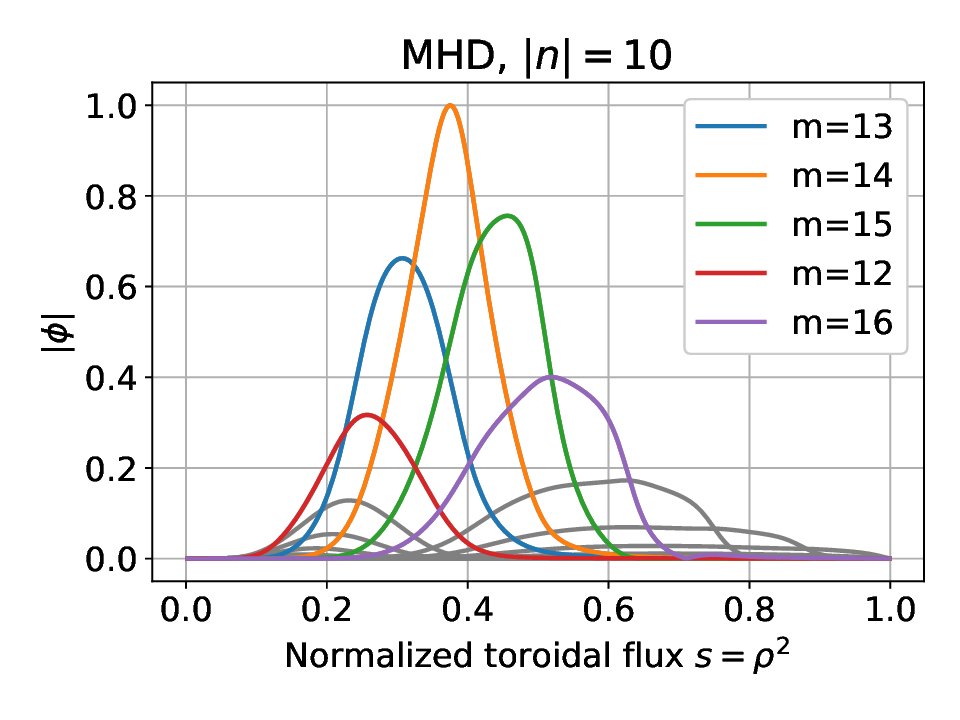}
        \caption{}
    \end{subfigure}
    \begin{subfigure}{0.32\linewidth}
        \includegraphics[width=\linewidth]{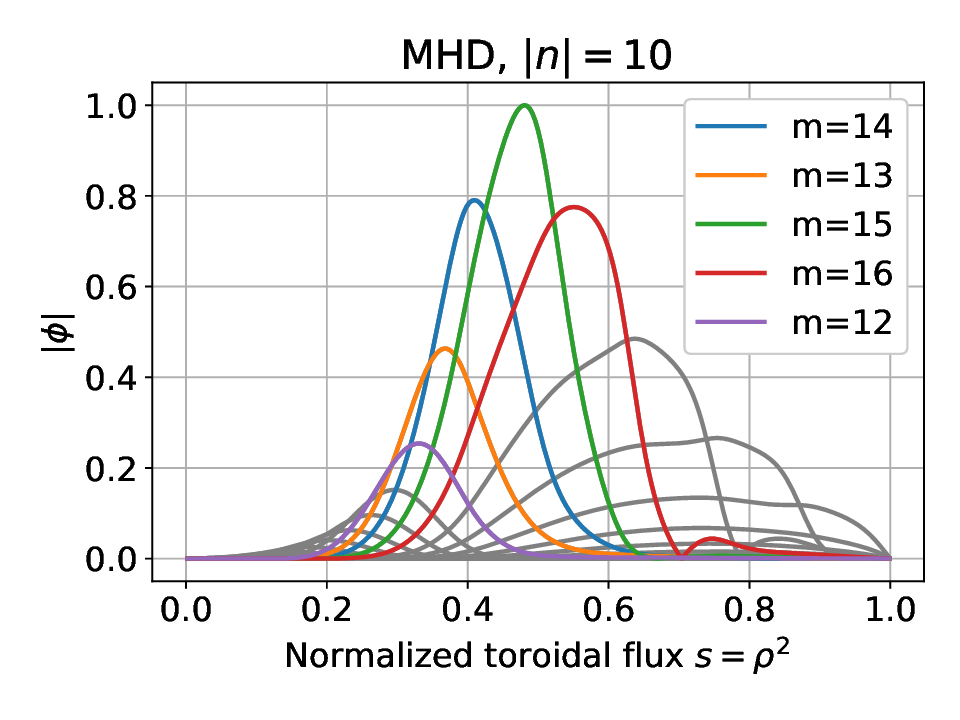}
        \caption{}
    \end{subfigure}
    \hfill    
    \begin{subfigure}{0.32\linewidth}
        \includegraphics[width=\linewidth]{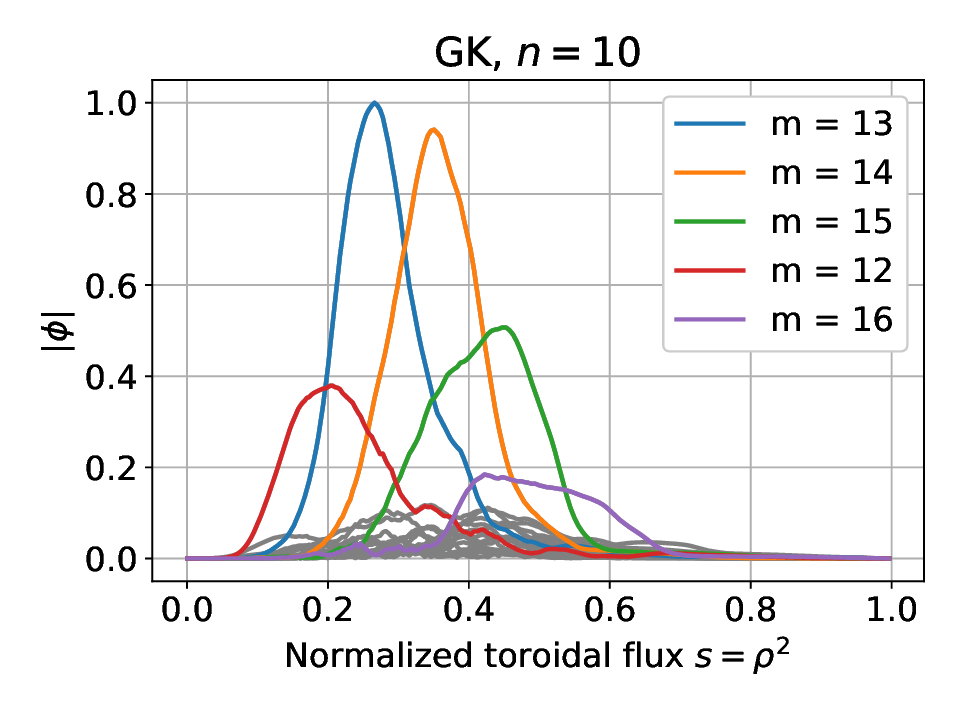}
        \caption{}
    \end{subfigure}
    \begin{subfigure}{0.32\linewidth}
        \includegraphics[width=\linewidth]{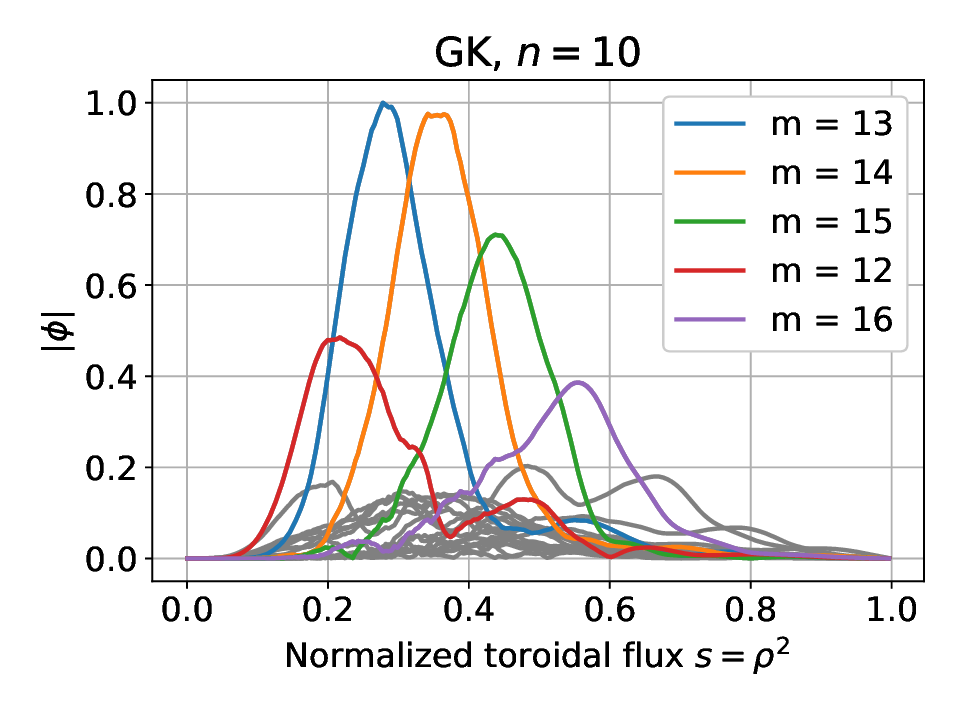}
        \caption{}
    \end{subfigure}
    \begin{subfigure}{0.32\linewidth}
        \includegraphics[width=\linewidth]{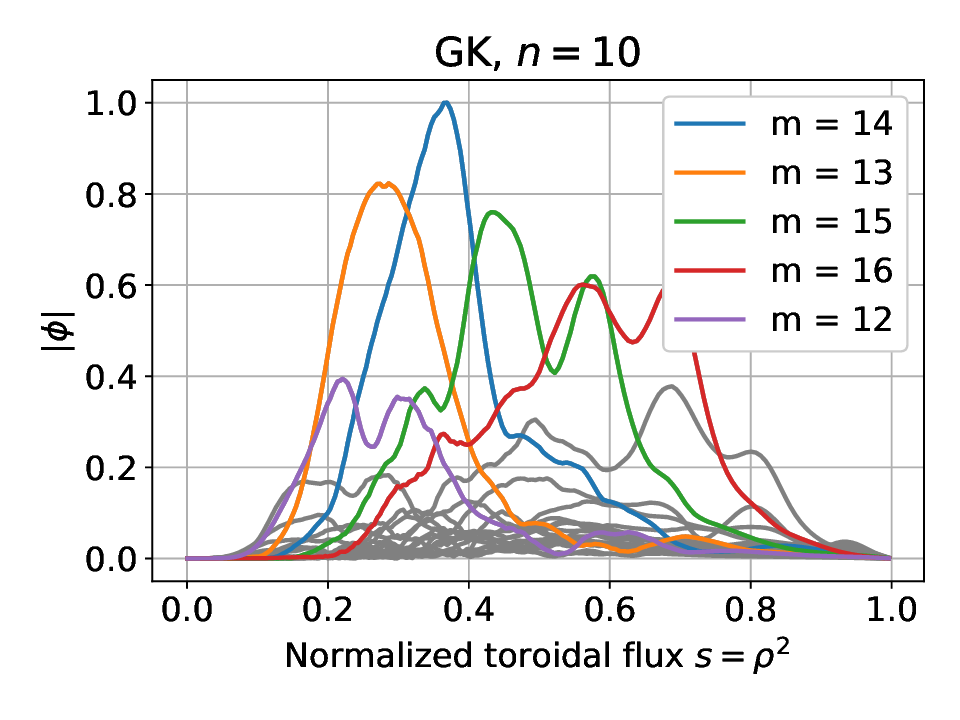}
        \caption{}
    \end{subfigure}
    \caption{Mode structures of the normalized electrostatic potential $|\phi|$ for the $\eta=1$ case. (a) - (c) show CAS3D results and (d) - (f) display EUTERPE results with gyro-average retained. Both codes display the following values $\left<\beta\right>=0.51\%$, $1.1\%$ and $2.2\%$, respectively from left to right column. The electrostatic potential of the MHD result is obtained by transforming the displacement $\xi$.}
    \label{fig:TokCircCAS3D_eta1_modes}
\end{figure}

Mode structures of the electrostatic potential obtained with EUTERPE and CAS3D are shown in \Fref{fig:TokCircCAS3D_eta1_modes}. Even when the gyro-average is retained, the two codes agree on the shape and position of the modes at low $\left<\beta\right>$. As $\left<\beta\right>$ increases, the similarity degrades, resulting in differing dominant harmonics and their positions. Nevertheless, the overall agreement across the entire $\left<\beta\right>$-scan remains satisfactory, confirming that the chosen parameters place the plasma in a fluid-like regime.\\

\begin{figure}
    \centering
    \begin{subfigure}{0.49\linewidth}
        \includegraphics[width=\linewidth]{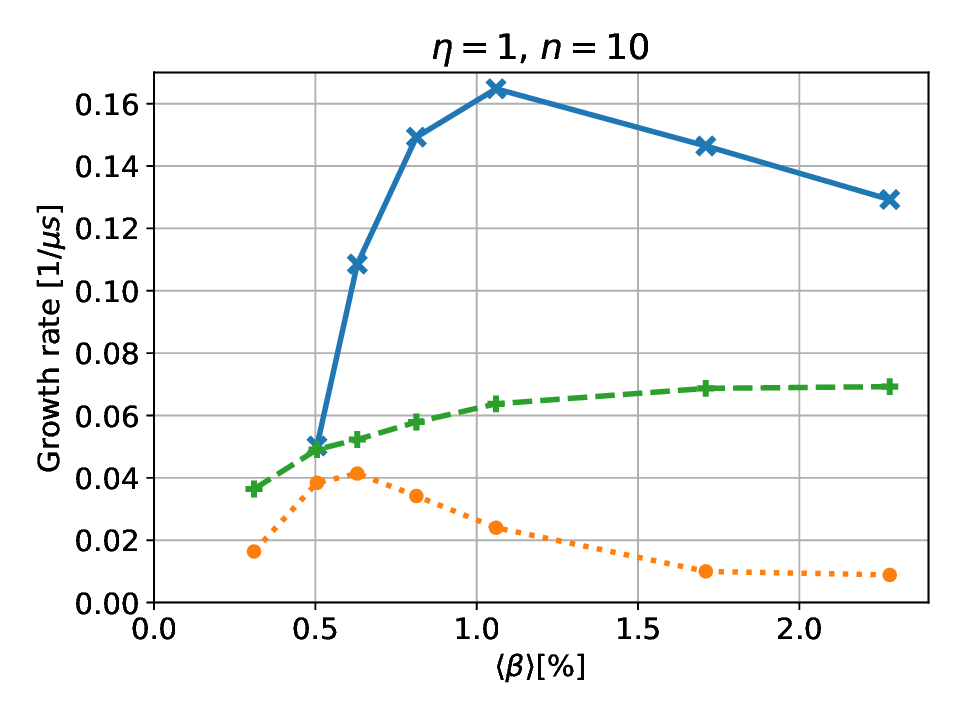}
        \caption{}
        \label{fig:TokCircCAS3D_eta1_gamma_temp}
    \end{subfigure}
    \begin{subfigure}{0.49\linewidth}
        \includegraphics[width=\linewidth]{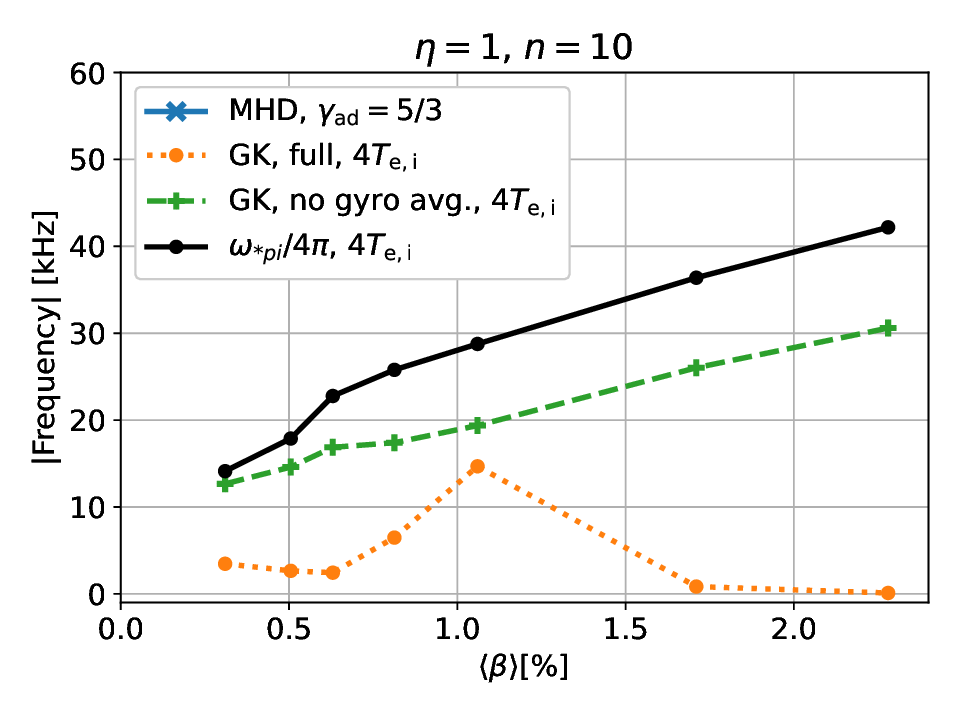}
        \caption{}
        \label{fig:TokCircCAS3D_eta1_freq_temp}
    \end{subfigure}
    \caption{Investigating the effect of an increased $\rho^*$ for the $\eta=1$ case from \Fref{fig:TokCircCAS3D_eta1_results}. Frequencies in (b) are shown as absolute values. The curve labeled ``$4T_\mathrm{e,i}$'' rotates into the electron diamagnetic direction. The legend in (b) also applies to (a), except for the frequency $\omega_\mathrm{*pi}$.}
    \label{fig:TokCircCAS3D_eta1_results_temp}
\end{figure}

When we change the equilibrium from the $\eta=6.67$ case to the $\eta=1$ case, temperatures were also reduced by a factor of $\approx 7$. Consequently, the normalized ion Larmor radius $\rho^*$ changes between the cases, and we must assess the impact of this change. To isolate the $\rho^*$ effect we increase both temperatures by a factor of 4 for the $\eta=1$ profiles and again solve the MHD and gyrokinetic eigenvalue problems using CAS3D and EUTERPE, respectively. The results are shown in \Fref{fig:TokCircCAS3D_eta1_results_temp}. Note that the density is used in the normalization of the MHD growth rates and the density has been adjusted accordingly so that the MHD growth rates are scaled correctly to the new temperatures. When the gyro-average is retained, raising $\left<\beta\right>$ above $0.8\%$ causes the growth rates to decrease, reproducing the trend seen in earlier scans.  The frequencies do not follow a clear systematic pattern. Conversely, when the gyro-average is disabled, modes appear that are resonant with the $q$-profile, just as in the $\eta=6.67$ case shown in \Fref{fig:TokCircCAS3D_EUmode_noFLR}. In this situation, both the growth rates and the frequencies increase with $\left<\beta\right>$. The frequency trend approximately follows the ion diamagnetic frequency $\omega_\mathrm{*pi}$. In \Sref{sec:discussion} we will show that these modes are produced by the coupling between a beta-induced Alfvénic eigenmode (BAE) and a KBM.

\subsection{\texorpdfstring{$\eta=0.5$}{\eta=0.5}-case}
\label{sec:circTok_CAS3D_eta05}

\begin{figure}
    \centering
    \begin{subfigure}{0.49\linewidth}
        \includegraphics[width=\linewidth]{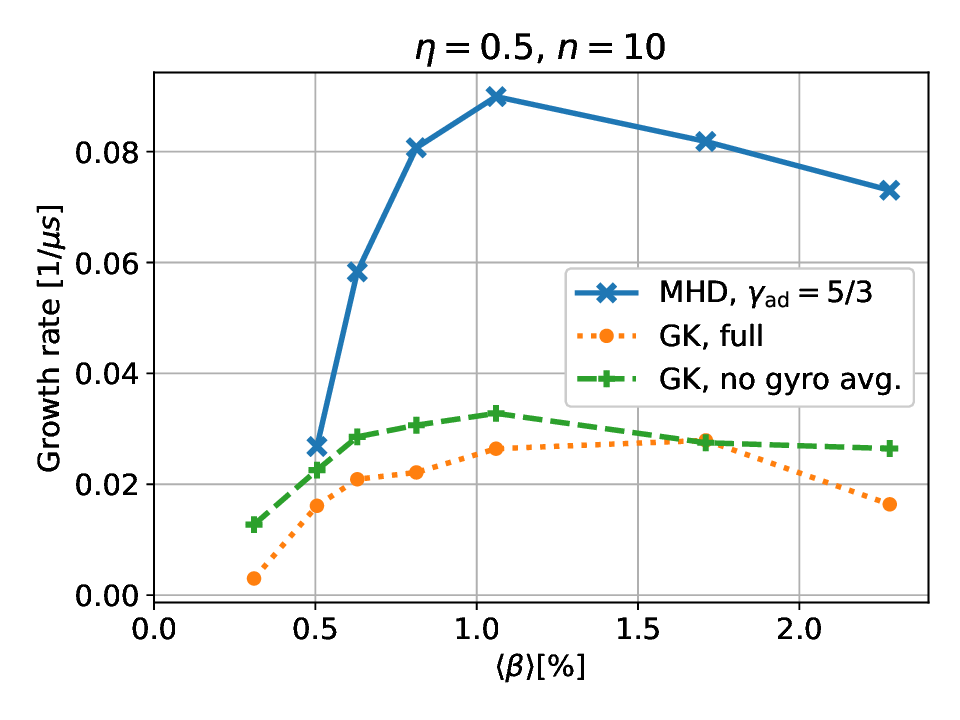}
        \caption{}
        \label{fig:TokCircCAS3D_eta0.5_gamma}
    \end{subfigure}
    \begin{subfigure}{0.49\linewidth}
        \includegraphics[width=\linewidth]{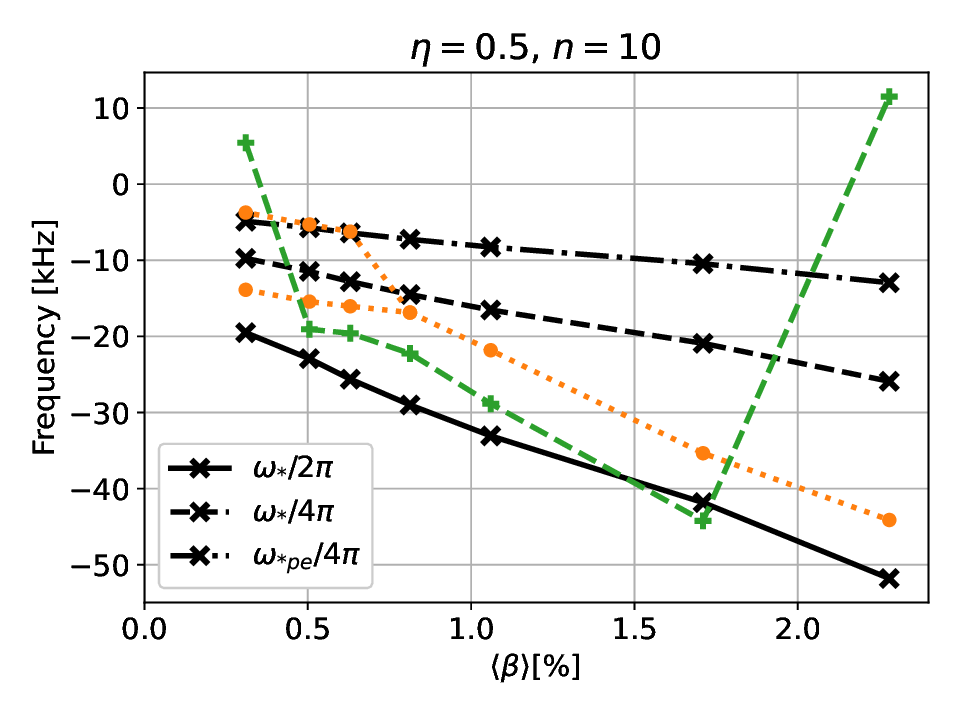}
        \caption{}
        \label{fig:TokCircCAS3D_eta0.5_freq}
    \end{subfigure}
    \caption{Comparison of growth rates and frequencies for $\eta=0.5$ between EUTERPE and CAS3D. A second CAS3D scan with $\gamma_\mathrm{ad}=10^{-5}$ results in slightly up-shifted growth rates, but is not displayed here for visual clarity. The legend of figure (a), excluding the MHD results, also applies to figure (b). For low values of $\beta$ in the gyrokinetic simulations without gyro-average, two clear and distinct frequencies could be observed and are shown in figure (b). When using full gyrokinetics, the lower frequency disappears.}
    \label{fig:TokCircCAS3D_eta0.5_results}
\end{figure}

Since a reduction from $\eta=6.67$ to $\eta=1$ already improved the agreement between the gyrokinetic and the MHD model, it is natural to ask whether a further reduction to $\eta=0.5$ yields additional improvement. Accordingly, the profiles are changed from the $\eta=1$ case to $\eta=0.5$ as described in \Sref{sec:setup}.\\

\Fref{fig:TokCircCAS3D_eta0.5_results} compares the EUTERPE and CAS3D results. Although the qualitative trends calculated by CAS3D are reproduced, the gyrokinetic growth rates are lower and the overall agreement is poorer than for the $\eta=1$ case. For $\left<\beta\right><0.8\%$ the mode exhibits two distinct frequencies, and a single dominant frequency cannot be identified. Consequently both frequencies are plotted in \Fref{fig:TokCircCAS3D_eta0.5_freq}. The lower frequency follows $\omega_\mathrm{*pe}/2$ and corresponds to a KBM. The higher frequency lies between $\omega_\mathrm{*pe}$ and $\omega_\mathrm{*}$ and initially follows the BAE branch. This is discussed further in \Sref{sec:discussion}. For $\left<\beta\right>\geq0.8\%$ the mode structures and trends become similar to those observed for $\eta=1$.

When the gyro-average is disabled, both growth rates and frequencies increase slightly in magnitude. At both low and high $\left<\beta\right>$, we observe modes with positive frequency, which are likely KBMs because their frequency aligns closely with half the ion diamagnetic frequency (this frequency is omitted for visual clarity).\\

The mode structures (not shown) exhibit a behavior comparable to that seen for the $\eta=1$ case in \Fref{fig:TokCircCAS3D_eta1_modes} when the gyro-average is deactivated: at low and intermediate $\left<\beta\right>$ the two codes agree reasonably well, the agreement deteriorates at high $\left<\beta\right>$ where kinetic effects become important.\\

\section{Discussion of results}
\label{sec:discussion}
We have so far examined KBMs from two complementary perspectives: the first one being focused on micro-instabilities 
(i.e. the ITG-KBM transition) and the second one on MHD stability. 
Now, we will summarize and systematize our main findings referring to earlier results \cite{zoncaKineticTheoryLowfrequency1996,maTheoreticalStudiesLowfrequency2022}  
whereby we introduce a third perspective based on low-frequency Alfvén modes.\\

In \Sref{sec:circTok} we considered the standard micro-instabilities picture. Using a vacuum magnetic equilibrium we reproduced the well-known ITG-to-KBM transition, but the transition disappeared when we instead used pressure-consistent equilibria. Apparently, equilibrium effects stabilize the KBM: the growth rates are reduced and the critical $\beta$ for KBMs is shifted upward as the equilibrium pressure $\beta_\mathrm{EQ}$ increases. To the best of our knowledge, similar investigations have only been carried out in a local flux-tube model by Niiro et al. \cite{niiroPlasmaBetaDependence2023} and Aleynikova et al. \cite{aleynikovaKineticBallooningModes2018}. In the latter work, a stability map for $\beta$ and $k_y\rho_\mathrm{s}$ was presented for both consistent and inconsistent equilibria. The map shows that the KBM becomes much more benign and almost vanishes for the consistent case. 
The effect of using a consistent equilibrium instead of a vacuum equilibrium for a typical ITG-to-KBM transition was also investigated by Mishchenko et al. \cite{mishchenkoGlobalGyrokineticSimulations2023} using a global code, but none of these works has received broad attention in the community.\\

In \Sref{sec:circTok_CAS3D} we investigated KBMs from the MHD perspective. With the full gyrokinetic model we could only recover a subset of the MHD-unstable modes. The outcomes were highly sensitive to the chosen physical parameters, especially the ratio of temperature gradient to density gradient. Substantial diamagnetic stabilization is evident: the agreement between the MHD and gyrokinetic models depends on whether gyro-averaging is retained or not. Better agreement is generally found when the gyro-average is omitted, but significant kinetic effects remain even without gyro-averaging, leading to sizable discrepancies between the two models. This stresses the importance of considering kinetic effects.\\

A third viewpoint can provide additional insight on KBMs: low-frequency Alfvén modes, in which the shear Alfvén continuum couples to the sound continuum. Representative studies are those of Zonca et al. \cite{zoncaKineticTheoryLowfrequency1996} and Ma et al. \cite{maTheoreticalStudiesLowfrequency2022}, where KBMs and beta-induced Alfvén eigenmodes (BAE) are treated with diamagnetic effects and core-plasma ion compressibility. In the limit of $\omega\sim\omega_\mathrm{*pi} \sim \omega_\mathrm{thi,trans}$ (with the ion transit frequency $\omega_\mathrm{thi,trans}=v_\mathrm{th,i}/(qR_0)=\sqrt{2k_\mathrm{B}T_\mathrm{i}/m_\mathrm{i}}/qR_0$), the KBM and BAE branches become strongly coupled via ion compressibility. These two branches are independent only for $\eta=0$ \cite{zoncaKineticTheoryLowfrequency1996}. Otherwise they are coupled, so a pure KBM branch is the exception rather than the rule. Note that, in another context, the KBM/BAE coupling has been shown to be important in \cite{bierwageGyrokineticAnalysisLown2017}.\\

\begin{figure}
    \centering
    \begin{subfigure}{0.49\linewidth}
        \includegraphics[width=\linewidth]{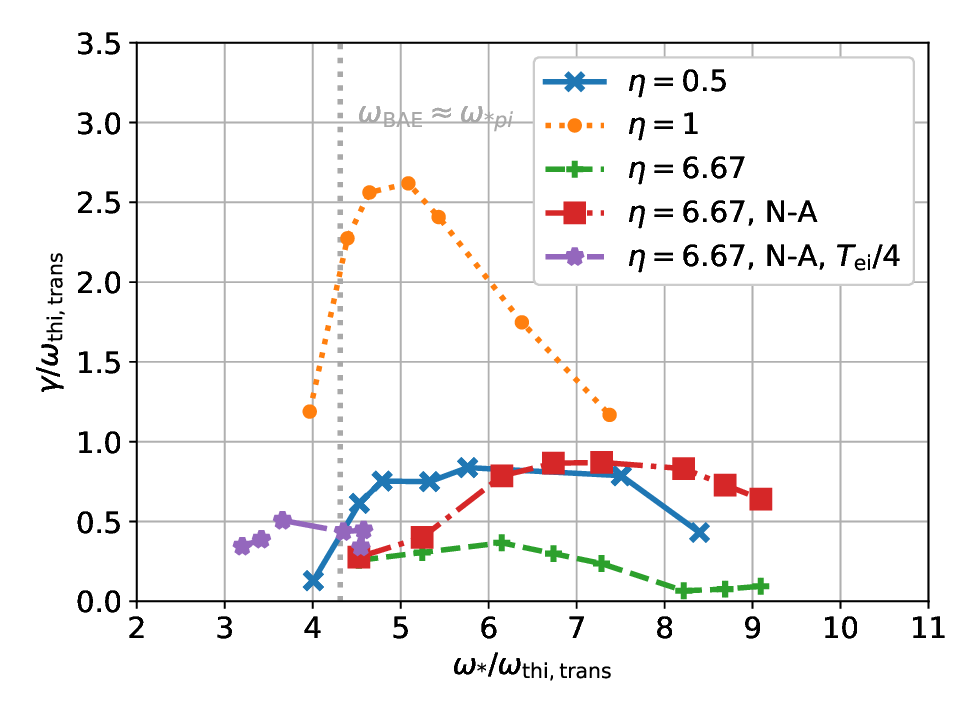}
        \caption{}
        \label{fig:TokCircCAS3D_theory_gamma2}
    \end{subfigure}
    \begin{subfigure}{0.49\linewidth}
        \includegraphics[width=\linewidth]{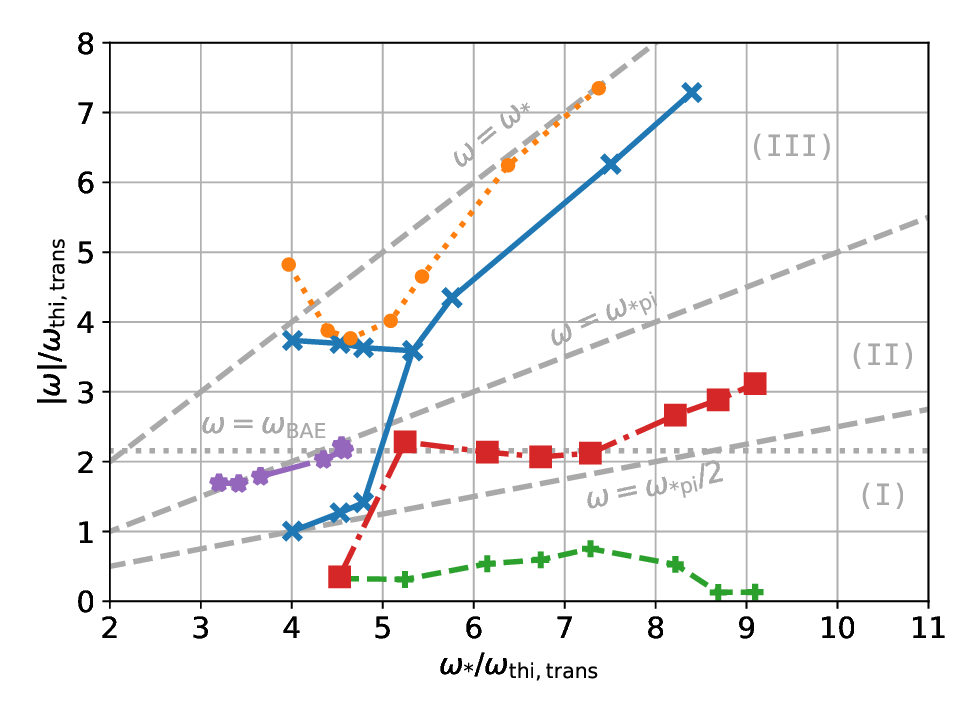}
        \caption{}
        \label{fig:TokCircCAS3D_theory_freq2}
    \end{subfigure}
    \caption{Normalized (a) growth rates and (b) frequencies of EUTERPE results from \Sref{sec:circTok_CAS3D} are displayed against $\omega_\mathrm{*}/\omega_\mathrm{thi,trans}$. Growth rates and frequencies are normalized by the thermal ion transit frequency $\omega_\mathrm{thi,trans}$. For all frequencies from the simulations as well as diamagnetic frequencies the modulus is taken. The ``N-A'' label indicates simulations without gyro-average. The grey indicated frequencies sort the resulting frequencies into three groups ((I), (II) and (III)). These are explained in the text.}
    \label{fig:TokCircCAS3D_theory}
\end{figure}

In the spirit of Refs.~\cite{zoncaKineticTheoryLowfrequency1996,maTheoreticalStudiesLowfrequency2022}, the frequencies and growth rates from \Sref{sec:circTok_CAS3D} have been normalized to $\omega_\mathrm{thi,trans}$ and plotted against $\omega_\mathrm{*} / \omega_\mathrm{thi,trans}$ in \Fref{fig:TokCircCAS3D_theory}. The additional reference frequencies $\omega_\mathrm{*}$, $\omega_\mathrm{*pi}$, $\omega_\mathrm{*pi}/2$ and the BAE frequency $\omega_\mathrm{BAE}=\omega_\mathrm{thi,trans}q\sqrt{7/4+T_\mathrm{e}/T_\mathrm{i}}$ are also indicated (caused by our parameter choice, $\omega_\mathrm{*}/2=|\omega_\mathrm{*pi}|=|\omega_\mathrm{*pe}|$).\\

\Fref{fig:TokCircCAS3D_theory_gamma2} shows that the growth rates exhibit a pronounced peak for the $\eta=1$ case. According to Zonca et al. \cite{zoncaKineticTheoryLowfrequency1996}, a sharp peak is expected when the BAE and KBM branches are strongly coupled, i.e. when the BAE frequency and the ion diamagnetic frequency are comparable ($\omega_\mathrm{*pi}\approx\omega_\mathrm{BAE}$). This condition is marked by the grey vertical line and coincides with the observed peak, suggesting that the $\eta=1$ case corresponds to a coupled BAE/KBM branch.

\Fref{fig:TokCircCAS3D_theory_freq2} reveals a clear separation of the cases into three groups:
\renewcommand{\theenumi}{\Roman{enumi}}%
\begin{enumerate}
    \item ``Kinetic modes'' which cannot be explained by MHD with the frequencies $\omega\lesssim\omega_\mathrm{*pi}/2$.
    \item ``Diamagnetic stabilization'': strong effect of the gyro-average on the stability with results approaching MHD if the gyro-average is omitted in the gyrokinetic equation. In this case, the mode frequencies are in the interval $\omega_\mathrm{*pi}/2\lesssim\omega\lesssim\omega_\mathrm{*pi}$. 
    \item MHD-like instabilities ($\eta=0.5$ and $\eta=1$) with the frequencies $\omega_\mathrm{*pi}<\omega\lesssim\omega_\mathrm{*}$. 
\end{enumerate}

The first regime is represented by the $\eta=6.67$ case. Strong kinetic effects suppress the growth rates and push the frequencies into the acoustic regime. When the gyro-average is switched off, the frequencies move onto the BAE-KBM branch: the mode first follows the BAE frequency and then becomes diamagnetic, i.e. it turns into a KBM. Reducing $\rho^*$ via the temperature shifts the branch into a pure KBM.\\

For the $\eta=0.5$ case two frequencies appear at low $\beta$. The lower one matches $\omega_\mathrm{*pi}/2$ and therefore belongs to the KBM range. Deactivating the gyro-average makes this lower branch disappear, indicating that it is diamagnetically destabilized. Ignoring this branch, the remaining frequencies for the $\eta=0.5$ and $\eta=1$ cases are almost constant for $\omega_\mathrm{*} / \omega_\mathrm{thi,trans} \lesssim 5$. They are also roughly double the BAE frequency, suggesting a higher-harmonic BAE with all kinetic effects retained. Deactivating the gyro-average raises both growth rates and frequencies slightly, bringing them even closer to $2 \, \omega_\mathrm{BAE}$ and confirming the kinetic modification of the BAE branch. For $\omega_\mathrm{*} / \omega_\mathrm{thi,trans}\gtrsim5$ the frequencies scale as $\omega\sim\omega_\mathrm{*}$, which is consistent with a KBM-type mode.\\

\section{Summary}
Using a global gyrokinetic approach, we have investigated KBMs from three distinct perspectives: micro-instabilities (the typical ITG-to-KBM transition), MHD stability and lastly low-frequency Alfvén modes and the shear Alfvén continuum (examples for all three perspectives can be found in Refs.~\cite{nuhrenbergGyrokineticSimulationsMagnetohydrodynamic2025,maTheoreticalStudiesLowfrequency2022,aleynikovaInfluenceMagneticConfiguration2022}). 
With respect to micro-instabilities, we have confirmed that the ITG-KBM transition can disappear entirely when pressure-consistent equilibria are employed. 
In the parameter range relevant to MHD, we demonstrate that gyrokinetics can recover MHD-unstable modes although kinetic effects may become important at low frequencies and spoil agreement for realistic conditions.
Diamagnetic effects may often account for the differences but other kinetic contributions such as the wave-particle resonances can contribute as well. 
Finally, we revealed a parameter regime where a rich spectrum of hybrid low-frequency Alfvén modes appears when transitioning into the higher-beta regime.

In summary, a pure KBM or a pure MHD mode (the usual objects of investigations) are idealizations which normally cannot be isolated for realistic parameters because of the strong couplings (between different waves or between the waves and the particles) present in the system. 
Nevertheless, this is an encouraging result because the KBMs need not be as detrimental as suggested, e.g. in Ref.~\cite{mulhollandFinitevTurbulenceWendelstein2024}. 
Likewise, the Large Helical Device stellarator has operated successfully in MHD unstable regimes without major MHD crashes \cite{sakakibaraStudyMHDStability2017}. 
Our results provide some insight into why MHD modes may not dominate even when the equilibrium is formally MHD-unstable. 
Further work is required to understand in which parameter regimes MHD models provide a reasonable approximation for high $\left<\beta\right>$-physics and to assess whether KBMs are generally benign in global simulations. Similarly, there is still the open question regarding the complex behavior of the nonlinear evolution of KBMs when considering the global approach.\\

\section*{Acknowledgement}
The authors thank Fulvio Zonca, Zhisong Qu, Xavier Garbet and Yanick Sarazin for helpful discussions. The authors also thank Alessandro Zocco for useful conversations during the course of this work and highlighting strengths of this work. Additional thanks are given to Per Helander for helping with editing this paper.
The authors gratefully acknowledge RES resources provided by BSC in 
MareNostrum5 to FI-2025-2-0040 and FI-2025-3-0048.

\section*{Funding}
This work has been carried out within the framework of the EUROfusion Consortium, funded by the European Union via the Euratom Research and Training Programme (Grant Agreement No 101052200 — EUROfusion). Views and opinions expressed are however those of the author(s) only and do not necessarily reflect those of the European Union or the European Commission. Neither the European Union nor the European Commission can be held responsible for them.

\bibliographystyle{unsrt}
\bibliography{literature}

\end{document}